\documentclass[10pt,final]{elsart}
\usepackage{amsmath,amssymb}
\usepackage[dvips]{graphicx}
\usepackage{color}

\newcommand{\fm}{\;\text{fm}}
\newcommand{\GeV}{\;\text{GeV}}

\newcommand{\tr}{\mathrm{tr}}
\newcommand{\rmi}{\mathrm{i}}
\newcommand{\rmd}{\mathrm{d}}
\newcommand{\rme}{\mathrm{e}}
\newcommand{\xt}{\boldsymbol{x}_\perp}
\newcommand{\kt}{\boldsymbol{k}_\perp}

\newcommand{\dtau}{\Delta\tau}

\begin{document}
\begin{frontmatter}

\title{The evolving Glasma}

\author[Keio]{Kenji Fukushima}
and
\author[Saclay]{Fran\c{c}ois Gelis}

\address[Keio]{Department of Physics, Keio University,
               Kanagawa 223-8522, Japan}
\address[Saclay]{Institut de Physique Th\'{e}orique
         (URA 2306 du CNRS)
         CEA/DSM/Saclay,\\
         91191, Gif-sur-Yvette Cedex, France}

\begin{abstract}
  We extensively study the growing behavior of the energy and the
  pressure components depending on the space-time rapidity in the
  framework of the Glasma, which describes the early-time dynamics in
  the ultra-relativistic heavy-ion collisions.  We simulate the Glasma
  solving the classical equations of motion in the SU(2) Yang-Mills
  theory and systematically investigate the dependence of the Glasma
  instability on the model parameters.  We have checked that the
  transverse and longitudinal grid sizes in our simulation are large
  enough to handle cutoff effects under control.  By comparing the
  numerical results from several initial conditions with different
  magnitudes of instability seed and also those with different
  wave-numbers for rapidity fluctuations, we clearly see that unstable
  modes dominantly grow up in the linear regime and we also confirm
  non-linear effects in the time evolution.  To extract more detailed
  information on the evolving Glasma, we decompose the energy into the
  components in terms of rapidity wave-numbers.  We observe an energy
  flow from low wave-number modes into higher wave-number modes due to
  non-linearity in the equations of motion.  We find that the energy
  spectrum approaches an asymptotic scaling that is consistent with
  Kolmogorov's power-law form even in the expanding system of the
  Glasma.
\end{abstract}
\begin{keyword}
Color Glass Condensate, Glasma, Relativistic heavy-ion collision,
Instability, Kolmogorov spectrum
\end{keyword}
\end{frontmatter}


\section{Introduction}

The formation of a quark-gluon plasma (QGP) is achieved in
ultra-relativistic heavy-ion collision experiments at Relativistic
Heavy-Ion Collider (RHIC) in BNL and at the Large Hadron Collider
(LHC) in CERN.\ \ More and more information on properties of strongly
interacting quark-gluon matter in quantum chromodynamics (QCD) is
being accumulated.  However, despite a number of interesting
discoveries and new notions established in experiments, the
thermalization mechanism that leads to the QGP still lacks a deep
understanding based on QCD first principles.

There have been many attempts to estimate the thermalization time
scale, that is phenomenologically known to be
$\tau_{\rm eq}\lesssim 1\fm/c$ from analysis in hydrodynamic
models~\cite{Kolb:2002ve,Heinz:2004pj}.  It was
argued~\cite{Baier:2000sb} that perturbative QCD processes lead to a
thermalization time of order $\alpha^{-13/5}\,Q_s^{-1}$ where the
saturation momentum $Q_s$ is a characteristic hard scale in the
high-energy reactions; $Q_s(x\sim p_\perp/\sqrt{s})= 1$-$2\GeV$ in the
typical RHIC kinematics.  If the prefactor is of order unity, this
estimate would yield a thermalization time around several
$\text{fm}/c$, which is too slow to account for the RHIC observations.
It has also been suggested that QCD perturbation theory, no matter at
which order one truncates it, may be unable to describe thermalization
at all~\cite{Kovchegov:2005ss}.  Plasma instabilities, particularly
the Weibel
instability~\cite{Arnold:2003rq,Arnold:2004ti,Arnold:2005vb,%
Arnold:2005ef,Randrup:2003cw,Mrowczynski:2005ki,Dumitru:2006pz,%
Berges:2008zt}, could be promising candidates for the early
thermalization mechanism~\cite{Bodeker:2005nv} and have been well
investigated in the hard-loop
formalism~\cite{Rebhan:2004ur,Rebhan:2005re,Romatschke:2006wg,%
Rebhan:2008uj,Ipp:2010uy}.  There are also various other
approaches~\cite{Muller:1992iw,Biro:1994sh,Kunihiro:2010tg,Nishiyama:2010mn}
including a framework based on the AdS/CFT
correspondence~\cite{Chesler:2008hg,Chesler:2009cy,Kovchegov:2009du,%
Balasubramanian:2010ce}, in which thermalization appears as the
formation of a black hole horizon in the dual theory.

Parallel to the analysis on plasma instabilities, an important
observation was made in a related theoretical framework.  The
occupation number of gluons at high energy is so huge that color field
amplitudes $A^\mu\sim \alpha_s^{-1/2}$ can be realized, and then
non-linearities must be taken into account to all orders.  Instead of
a plain perturbative expansion, then, an approximation in terms of
classical field equations becomes more suitable because it effectively
makes a resummation of large-amplitude fields $A_\mu$.  This
non-linear regime, whose natural time scale is of order
$Q_s^{-1}\simeq 0.2\fm/c$ at RHIC, is expected to be prevalent during
the early stages in heavy-ion collisions.  The QCD framework known as
the Color Glass Condensate (CGC)~\cite{Gelis:2007kn,Gelis:2010nm}
rearranges the perturbative expansion in terms of classical fields
(that correspond to the resummation of infinite sets of graphs), and
perturbations on top of these classical fields are of higher order in
$\alpha_s$.  The CGC was first developed in order to deal with the
saturation effect in the gluon distribution
function~\cite{McLerran:1993ni,McLerran:1993ka,McLerran:1998nk}.
Because we are interested in the early time ($\tau\lesssim Q_s^{-1}$)
dynamics before thermalization, the CGC description seems to be the
most suited for our purpose.  The initial condition just above the
forward light-cone (i.e.\ at $\tau=0^+$) for the classical field that
appears at lowest order in the CGC description can be nicely
formulated in the Bjorken (expanding)
coordinates~\cite{Kovner:1995ja,Kovner:1995ts}.  Because in the regime
of large occupation numbers, classical field theory and kinetic theory
essentially describe the same
dynamics~\cite{Son:1995wz,Mueller:2002gd}, it is conceivable that the
CGC can describe the isotropization and thermalization processes
qualitatively.  However, for this program to have a chance to work, it
is necessary to go beyond the lowest order description, since at
zeroth order it is well known that the pressure tensor never
isotropizes.

It is unfortunately impossible to obtain an analytical solution of the
classical Yang-Mills equations of motion for the heavy-ion collision
(two-source) problem, and therefore one has to resort to numerical
simulations.  In
Refs.~\cite{Krasnitz:1998ns,Krasnitz:1999wc,Krasnitz:2000gz},
heavy-ion collisions were simulated in the McLerran-Venugopalan (MV)
model (i.e.\ the CGC at lowest order, with a local Gaussian
distribution of color sources), formulated gauge-invariantly in terms
of link variables.  These numerical calculations were later confirmed
by an independent simulation in Ref.~\cite{Lappi:2003bi}.  The
peculiar feature in the CGC description at leading order is that the
initial condition for the classical fields is independent of the
space-time rapidity variable $\eta$, and the equations of motion
maintain $\eta$-independence during the time evolution.  Therefore,
the space-time evolution is entirely independent of $\eta$, that is,
invariant under boosts in the direction of the collision axis.
Intuitively, one can expect a longitudinally extended structure of the
initial fields, which is indeed a characteristic property of the
so-called ``Glasma'' state~\cite{Lappi:2006fp}, which could be a
source for the so-called ridge structure seen in the two-particle
correlations at RHIC and LHC~\cite{Dumitru:2008wn,Dusling:2009ni}.

Within the CGC framework, an instability has been found when the boost
invariant Glasma fields are
disturbed~\cite{Romatschke:2005pm,Romatschke:2006nk}.  This
instability has some similarities to the Weibel instability in
anisotropic plasmas, but the origin of the instability is not yet
fully clarified, though several interpretations have been
proposed~\cite{Fukushima:2007ja,Fukushima:2007yk,Fujii:2008dd,%
Fujii:2009kb,Dusling:2010rm}.
Also, it is still an issue how to properly formulate disturbances to
the boost invariant CGC background.  However, it is clear that such
perturbations arise when one considers higher order corrections to the
leading CGC picture~\cite{Fukushima:2006ax}, and therefore they should
not be ignored since they can alter phenomena such as particle
production~\cite{Gelis:2006yv,Gelis:2006cr,Gelis:2007pw},
etc.  Moreover, without including these perturbations, one cannot
quantify the thermalization processes.

One of the reasons why the origin of the Glasma instability lacks a
clear understanding is that the available numerical data from earlier
works~\cite{Romatschke:2005pm,Romatschke:2006nk} is rather limited.
This motivates us to carry out a more systematic survey of the Glasma
instability, exploiting a wider range of simulation parameters.  In
this work we will specifically investigate the dependence of the
Glasma instability on the following parameters:

\noindent {\bf[i]} \textit{Transverse grid size $N$:}~ It is known that
the initial energy density at $\tau=0^+$ is sensitive to the
ultraviolet and infrared regulators in the transverse direction, and
this sensitivity becomes milder at finite
$\tau$~\cite{Kovchegov:2005ss,Fukushima:2007ja,Lappi:2006hq}.  It is
therefore important to check whether continuum-limit results can be
obtained reliably by looking systematically at the dependence on the
transverse lattice spacing $a\equiv L/N$, where $L$ is the length of
the box and $N$ the number of lattice points in the transverse
direction.

\noindent {\bf[ii]} \textit{Longitudinal grid size $N_\eta$:}~
Physical results should also not be contaminated by ultraviolet
singularities in the longitudinal direction.  To avoid this, we will
use initial fluctuations with sufficiently small longitudinal
wave-number $\nu$ ($\nu$ is the Fourier conjugate to the rapidity
$\eta$).  We will confirm that there is no dependence on the
longitudinal grid size $N_\eta$ or the longitudinal lattice spacing
$a_\eta\equiv L_\eta/N_\eta$, at least until the longitudinal spectrum
extends into the ultraviolet regions at very late time.

\noindent {\bf[iii]} \textit{Magnitude $\Delta$ of the seeds:}~ In
general the strength of the instability increases with the magnitude
$\Delta$ of the initial $\eta$-dependent perturbations.  This
dependence is found to be very simple until the unstable modes become
large enough to be self-interacting: at moderate times, the
finite-$\nu$ modes obey the linearized dynamics and their amplitudes
grow linearly as $\Delta$ increases.

\noindent {\bf[iv]} \textit{Initial wave-number $\nu_0$:}~ The time
evolution depends on the initial condition in the longitudinal
direction, and in particular on the wave-number $\nu_0$ of the seeds.
We will show that the instability produced by the smallest
wave-number\footnote{In our conventions, the wave-number $\nu$ is
  defined to be an integer between $-N_\eta/2$ and $+N_\eta/2$.  The
  lowest non-zero wave-number is thus $\nu=\pm1$.}
($\nu_0=\pm1$) is the fastest and strongest in a wide range of
times.  We will also confirm the dominance of linearity and the
appreciable presence of residual non-linear effects by verifying that
the result from an initial condition with multiple $\nu_0$'s can be
approximated as the superposition of the results from initial
conditions with single wave-numbers $\nu_0$'s.

In this paper, we report on numerical results about these dependences
for the growing behavior at the late stage of the Glasma evolution.
These numerical results provide a pool of observations that we hope
will help to understand the microscopic mechanisms that drive the
Glasma instabilities.  In the section~\ref{sec:formulas}, we recall
the most important formulas for a lattice study of this problem.  In
the section~\ref{sec:BI}, we briefly expose some results in the purely
boost-invariant case.  These are well-known results already, and
merely serve as a check of our implementation.  In the
section~\ref{sec:notBI}, we describe what happens when the Glasma
fields are perturbed by a rapidity-dependent seed, and we investigate
thoroughly how the result depends on the parameters listed above.  We
explore the time evolution of the $\nu$-spectrum in the
section~\ref{sec:Lspectrum}.  Finally, the section~\ref{sec:summary}
is devoted to a summary and outlook.


\section{Lattice formulation of the problem}
\label{sec:formulas}

In this section, we summarize our equations, with emphasis on the
lattice specificities.  The main goal of this section is to avoid
confusions due to the existence of different conventions in the
literature for some quantities.

The link variables are defined as
$U_\mu(x)\equiv\rme^{-\rmi g a A_\mu(x)}$ with the gauge potential
$A_\mu(x)$ that connects the neighboring sites from $x$ and
$x+\hat{\mu}$.  This expression for the link variable is the most
important one in order to establish the correspondence between the
continuum and the lattice (link-variable) formulations.  Under a gauge
transformation by $V(x)$ the link variables transform as
\begin{equation}
 U_\mu(x)\to V(x)U_\mu(x)V^\dagger(x+\hat{\mu}) \;,
\end{equation}
and the chromo-electric fields as
\begin{equation}
 E^i(x)\to V(x)E^i(x)V^\dagger(x) \;.
\end{equation}
The plaquette variable is then defined as
\begin{equation}
 U_{\mu\nu}(x) \equiv U_\mu(x)U_\nu(x+\hat{\mu})U_\mu^\dagger(x+\hat{\nu})
  U_\nu^\dagger(x) \approx \exp\bigl[-\rmi g a^2 F_{\mu\nu}(x)\bigr] \;,
\label{eq:approxF}
\end{equation}
where the last form is an approximation which gives rise to the field
strength $F_{\mu\nu}\equiv\partial_\mu A_\nu-\partial_\nu A_\mu -\rmi
g[A_\mu,A_\nu]$ of the continuum formulation.  This is a very useful
expression to study the continuum limit.  The lattice spacing should
be the longitudinal one, i.e.~$a_\eta$, which is distinct from the
transverse one $a$, if the longitudinal direction is involved in the
displacements $\hat{\mu}$ or $\hat{\nu}$ in the above formula.  The
canonical momenta in terms of link variables are expressed as
\begin{equation}
 \partial_\tau U_i(x) = \frac{-\rmi g}{\tau} \,
  E^i(x) U_i(x) \;,\qquad
 \partial_\tau U_\eta(x) = -\rmi ga_\eta\tau\,
  E^\eta(x) U_\eta(x) \;.
\end{equation}
Here we note that in the above we already use variables made
dimensionless by the transverse lattice spacing $a$, and thus $a$ will
never appear explicitly in our formula.  Because $\eta$ is a
dimensionless number, $a_\eta$ does not bring any mass dimension
unlike $a$.

We can also discretize the time in the above equations in order to
solve the time evolution numerically for $U_i(x)$ and $U_\eta(x)$;
\begin{align}
 U_i(\tau'') &= \exp\bigl[ -2\dtau\cdot \rmi g
  E^i(\tau') / \tau' \bigr] U_i(\tau) \;,\\
 U_\eta(\tau'') &= \exp\bigl[ -2\dtau\cdot \rmi g a_\eta
  \tau' E^\eta(\tau') \bigr] U_\eta(\tau) \;,
\end{align}
where $\tau'=\tau+\Delta\tau$ and $\tau''=\tau+2\Delta$. Note that we
have not written explicitly the position arguments in the quantities
that appear in these equations.  Here, the exponentiation of $E^\mu$
is important; otherwise the updated $U_\mu(\tau'')$ would not be a
unitary matrix.  The equations of motion are discretized in the same
manner as
\begin{equation}
\begin{split}
 E^i(\tau') & = E^i(\tau-\dtau) + 2\dtau
  \frac{\rmi}{2g a_\eta^2\tau}\Bigl[ U_{\eta i}(x)
  +U_{-\eta i}(x) - \text{(h.c.)} \Bigr]_\tau \\
 &\qquad + 2\dtau \frac{\rmi\tau}{2g}\,
  \sum_{j\neq i} \Bigl[ U_{ji}(x)+U_{-ji}(x) - \text{(h.c.)} \Bigr]_\tau \;,
\end{split}
\end{equation}
for the transverse components and
\begin{equation}
 E^\eta(\tau') = E^\eta(\tau-\dtau) +
  2\dtau \frac{\rmi}{2g a_\eta\tau}\sum_{j=x,y} \Bigl[
  U_{j\eta}(x) + U_{-j\eta}(x) - \text{(h.c.)} \Bigr]_\tau \;,
\end{equation}
for the longitudinal component.  We recall again that all variables
including $\tau$ and $\dtau$ above are dimensionless, and expressed in
unit of the transverse spacing $a$.  One can make sure that these
equations are equivalent to the ordinary equations of motion in the
continuum limit, by using the approximate form~\eqref{eq:approxF}.

The initial conditions are derived from the requirement of avoiding
the singularity at the collision point.  It is written in terms of
$U_i^{(m)}$, the classical solution of the MV model for a single color
source (the label $m=1,2$ indicates which of the two nuclei produces
the corresponding field). This single-source classical solution is a
transverse pure-gauge,
\begin{equation}
 U_i^{(m)}(\xt) = V^{(m)}(\xt)V^{(m)\dagger}(\xt+\Delta x_i) \;,
\label{eq:Uin}
\end{equation}
where the gauge rotation matrix $V^{(m)}$ can be written as
\begin{equation}
 V^{(m)\dagger}(\xt) = \exp\bigl[\,\rmi g\Lambda^{(m)}(\xt) \bigr] \;.
\label{eq:wilson}
\end{equation}
Here, $\Lambda^{(m)}$ is the solution of the Poisson equation,
$\partial_\perp^2\Lambda^{(m)}(\xt) = -\rho^{(m)}(\xt)$. In the MV
model, the random color source $\rho^{(m)}$ is Gaussian
distributed\footnote{A Gaussian distribution, albeit one with
  non-local correlations, is also an approximate solution of the
  JIMWLK evolution equation~\cite{Iancu:2002aq}.}, with a two-point
correlation
\begin{equation}
 \langle\rho^{(n)}(\xt)\rho^{(m)}(\xt')\rangle =\delta^{nm}\,g^2\mu^2
  \delta(\xt-\xt') \;.
\end{equation}
Higher-point correlation functions are obtained as products of
two-point contractions, via Wick's theorem.  In the MV model, $\mu$ is
the only dimensionful parameter and it is related to the saturation
momentum $Q_s$.  Then, the initial conditions
are~\cite{Krasnitz:1998ns};
\begin{align}
 & U_i = \bigl( U_i^{(1)}+U_i^{(2)} \bigr)
  \bigl( U_i^{(1)\dagger} + U_i^{(2)\dagger} \bigr)^{-1} \;,
\label{eq:initial_U}\\
 & E^\eta = \frac{-\rmi}{4g} \sum_{i=x,y}\Bigl[ \bigl( U_i-1 \bigr)
  \bigl( U_i^{(2)\dagger}-U_i^{(1)\dagger} \bigr) \notag\\
 &\qquad + \bigl( U_i^\dagger(x\!-\!\Delta x_i)-1\bigr)\bigl(
  U_i^{(2)\dagger}(x\!-\!\Delta x_i)
  -U_i^{(1)\dagger}(x\!-\!\Delta x_i) \bigr)
  - \text{(h.c.)}\Bigr] \;,
\label{eq:initial_E}
\end{align}
written here in the slightly simpler case of the SU(2) color group.
These discretized equations of motion and initial conditions
completely define the numerical MV model. Our conventions are mostly
consistent with those of previous works, except for the fact that the
field $A_\eta$ is not treated as an adjoint scalar-field variable but
instead we also describe it in terms of a link variable $U_\eta$.  We
note that our formulation is reduced to the conventional one with the
adjoint scalar-field variable if we take $a_\eta$ to be sufficiently
small.  In addition, since our interest lies primarily in the
qualitative behavior of the Glasma instability, we disregard the color
neutralization that has been imposed onto the MV model color sources
in other works~\cite{Krasnitz:2002mn,Krasnitz:2003jw}.

It is now straightforward to implement the numerical calculation for
gauge-invariant observables such as the components of the
energy-momentum tensor.  They are in the continuum convention written
down as
\begin{align}
 & \varepsilon = \bigl\langle T^{\tau\tau} \bigr\rangle
  = \bigl\langle\tr\bigl[ E_{_L}^2 + B_{_L}^2 + E_{_T}^2 + B_{_T}^2
    \bigr]\bigr\rangle \;,
\label{eq:E} \\
 & P_{_T} = \frac{1}{2}\bigl\langle T^{xx} + T^{yy} \bigr\rangle
  = \bigl\langle \tr\bigl[ E_{_L}^2 + B_{_L}^2 \bigr] \bigr\rangle \;,
\label{eq:P_{_T}} \\
 & P_{_L} = \bigl\langle \tau^2 T^{\eta\eta} \bigr\rangle
  = \bigl\langle\tr\bigl[ E_{_T}^2 + B_{_T}^2 - E_{_L}^2 - B_{_L}^2
    \bigr]\bigr\rangle \;,
\label{eq:P_{_L}}
\end{align}
where we define
\begin{equation}
 E_{_L}^2 \equiv E^{\eta a}E^{\eta a} \;,\qquad
 E_{_T}^2 \equiv \frac{1}{\tau^2}\bigl(E^{x a}E^{x a}+E^{y a}E^{y a}\bigr) \;.
\label{eq:E2}
\end{equation}
These formulas are implicitly summed over the color index $a$, for
$E_{_L}^2$ and $E_{_T}^2$ to be gauge invariant.  Here, the
chromo-magnetic field squared should be expressed in a way consistent
with Eq.~\eqref{eq:approxF}, that is,
\begin{equation}
 B_{_L}^2 = \frac{2}{g^2}\,\tr\bigl( 1-U_{xy} \bigr) \;,\qquad
 B_{_T}^2 = \frac{2}{(ga_\eta\tau)^2}\sum_{i=x,y}
  \tr\bigl( 1-U_{\eta i} \bigr) \;.
\label{eq:B2}
\end{equation}
These definitions are, however, not very convenient when we decompose
the energy into its Fourier components, as we will see later.  We
confirm that the energy-momentum tensor is traceless;
$T^\mu{}_{\mu}=T^{\tau\tau}-T^{xx}-T^{yy}-\tau^2 T^{\eta\eta}=0$.  It
should also be mentioned that our definition of $P_{_T}$ is different
from that in Refs.~\cite{Romatschke:2005pm,Romatschke:2006nk} by a
factor $2$ so that $P_{_T}=P_{_L}$ for an isotropic system.


\section{Boost Invariant Expansion}
\label{sec:BI}

We first check the consistency with the known results in the
boost-invariant case, i.e.\ in $(2+1)$ dimensions.  This is useful
also for later discussions on the spectral decomposition when we
consider $\eta$-dependent fluctuations.  In this section, we focus on
$\eta$-independent classical solutions, with initial conditions
specified in Eqs.~\eqref{eq:initial_U} and \eqref{eq:initial_E} that
are themselves independent of $\eta$.  Eventually, however, our goal
is to superimpose $\eta$-dependent fluctuations to these boost
invariant initial conditions in order to incorporate quantum
effects~\cite{Fukushima:2006ax}.

We note here that our numerical calculations are limited to the SU(2)
color group, so that we can get more samples and thus better
statistics for the same computational cost.  As discussed in
Ref.~\cite{Ipp:2010uy}, the SU(2) calculations capture the essential
properties of the Glasma instability, and we expect that the results
we report here for the SU(2) case would be qualitatively the same for
SU(3).


\subsection{Initial Configurations}

\begin{figure}
 \includegraphics[width=0.49\textwidth]{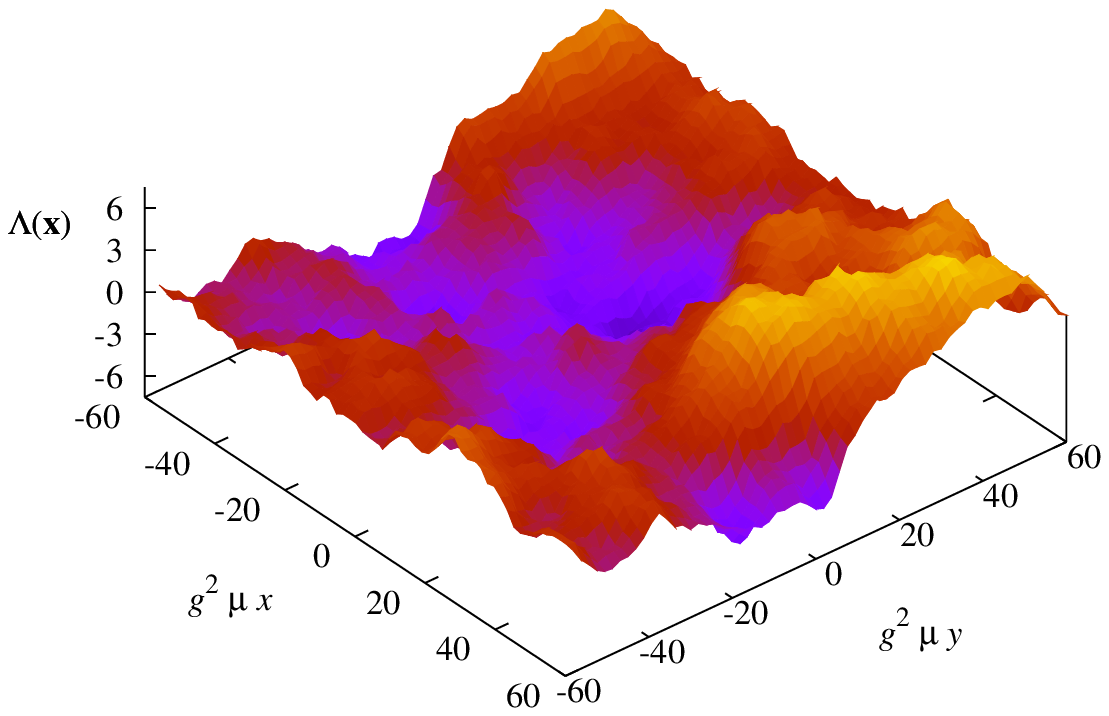}
 \includegraphics[width=0.49\textwidth]{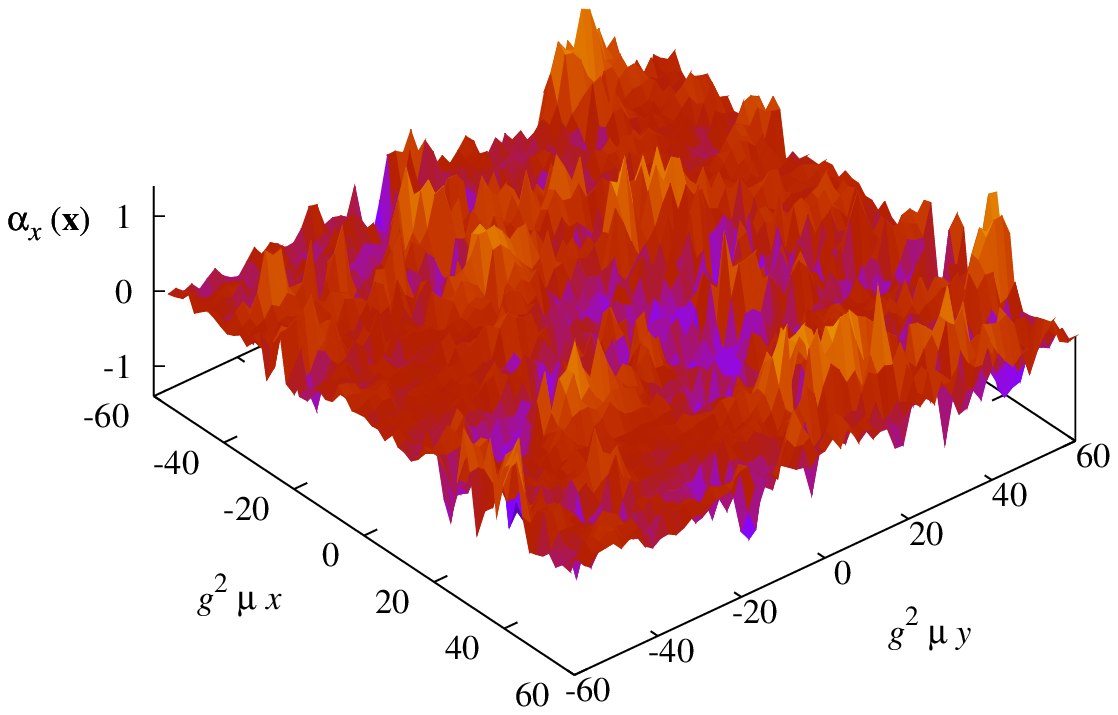}
 \caption{An initial configuration generated by a random color source
   $\rho(\xt)$.  The MV-model parameters are chosen as $N=64$ (number
   of transverse sites) and $g^2\mu L=120$.\ \ Left: The solution
   $\Lambda(\xt)$ of the Poisson equation.  Only the first color
   component $\Lambda^1(\xt)$ is plotted.\ \ Right: Corresponding
   gauge field $\alpha_x(\xt)$ calculated from
   $e^{-ig\Lambda(\xt)}e^{ig\Lambda(\xt+\hat{i})}=\exp[-\rmi
       g\alpha_i(\xt)]$.  Only the first color component
   $\alpha_x^1(\xt)$ is plotted.}
\label{fig:Lam}
\end{figure}

Let us first take a look at a typical fixed initial configuration,
without taking the ensemble average over the color sources, in order
to develop an intuitive picture for the Glasma state.  In the MV
model, the initial color distribution in the transverse plane is akin
to white noise, having no correlation between different transverse
sites\footnote{This property would be slightly altered if one imposed
  to the color distribution to be color neutral over patches
  corresponding to the size of a nucleon, or if one were using
  distributions of sources evolved with the JIMWLK equation -- in the
  latter case, it has been shown that the JIMWLK evolutions induces
  correlations among the sources over transverse distances of the
  order of $Q_s^{-1}$.}.  Since the solution of the Poisson equation
$\partial_\perp^2\Lambda^{(m)}(\xt)=-\rho^{(m)}(\xt)$ involves a
convolution with a massless (i.e.\ long ranged) transverse propagator,
$\Lambda(\xt)$ has a smooth structure as shown in the left panel of
Fig.~\ref{fig:Lam}.


Although $\Lambda(\xt)$ is fairly smooth due to the regularizing
effect of this convolution, the corresponding gauge field
$\alpha_i(\xt)$, which is a phase of $U_i(\xt)$ defined in
Eq.~\eqref{eq:Uin}, is far from smooth.  This is actually a
consequence of the pure-gauge form; contributions from globally smooth
parts in $\Lambda(\xt)$ cancel each other and the resulting
$\alpha_i(\xt)$ fluctuates strongly.  In other words, $\alpha_i(\xt)$
has a rough structure because a derivative of $\Lambda(\xt)$ undoes
the regularization provided by the convolution.  The $\alpha_i(\xt)$
shown in the figure \ref{fig:Lam} is the gauge field from a single
source only.  In a collision, the initial fields such as $E^\eta(\xt)$
and $B^\eta(\xt)$ involve a product of two $\alpha_i(\xt)$'s from two
independent sources, and thus have an even rougher structure.

It is worth mentioning that the ``color-flux tube'' structures, whose
transverse size is expected to be of order $Q_s^{-1}$, do not emerge
manifestly from the MV model since the source distribution has no
correlation length in the transverse direction.  This means that the
Nielsen-Olesen instability~\cite{Fujii:2009kb} is unlikely to occur at
least within the MV model.  Nevertheless, we expect that color-flux
tube structures would appear after the JIMWLK evolution from the
MV-model distribution.  In the mean-field approximation, in fact, the
transverse correlation function of color source has been already
studied in Ref.~\cite{Iancu:2002aq}, which should be taken into
account in the future works.  Also, to make a clear comparison with
the Weibel instability and the hard-loop estimates (especially early
works~\cite{Rebhan:2004ur,Rebhan:2005re}), it would be useful to
simplify the MV-model setup in (1+1) dimensions without transverse
dependences in such a way not to lose the qualitative features of the
Glasma instability that we will see in the section~\ref{sec:notBI}.
In this sense the results reported in the present paper should serve
as a foundation for future investigations in these directions.


\subsection{Time Evolution}
\label{sec:time_evolution}


\begin{figure}
 \includegraphics[width=0.49\textwidth]{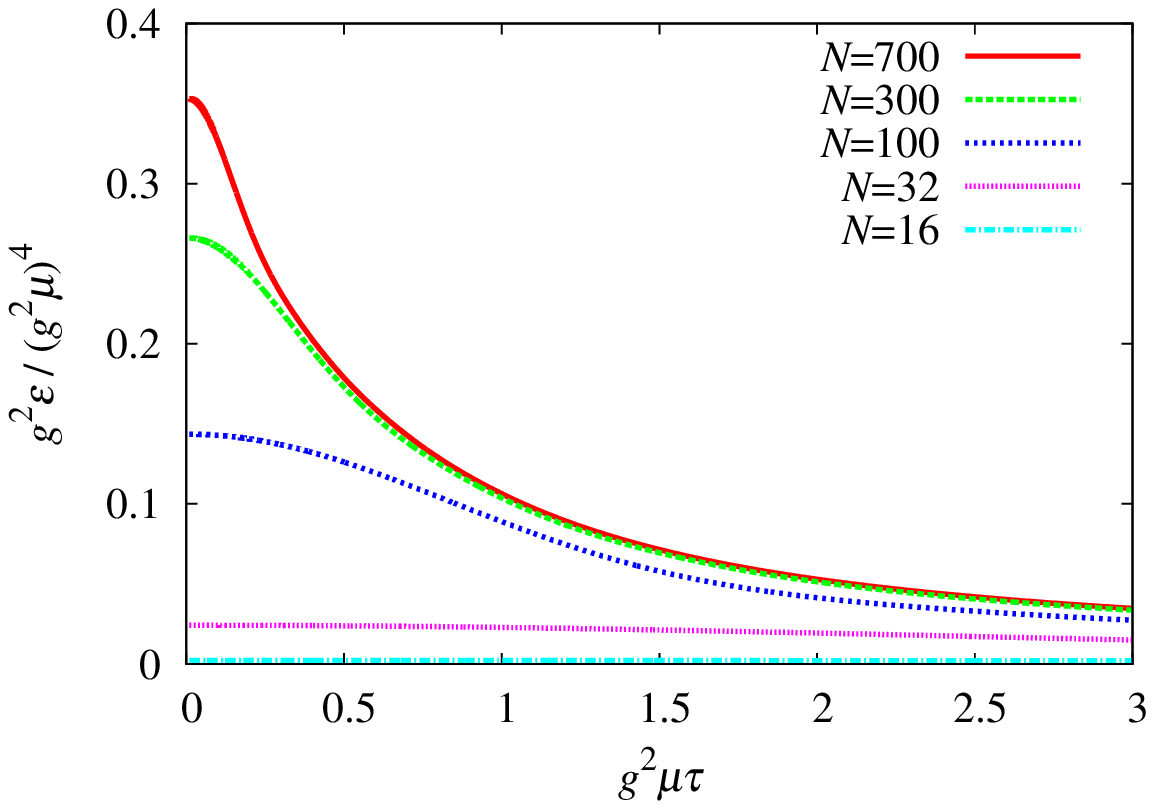}
 \includegraphics[width=0.49\textwidth]{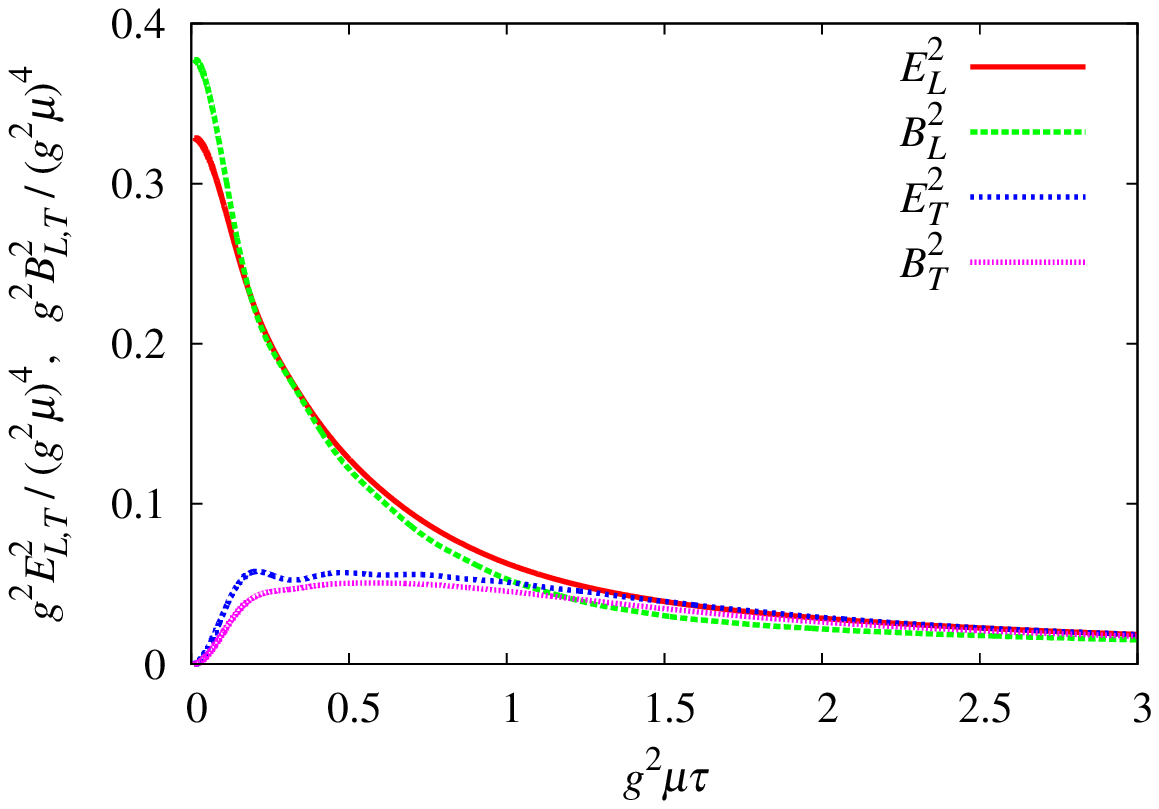}
 \caption{Left: Energy density~\eqref{eq:E} for various choices of the
   transverse grid size $N$.  An ensemble average is taken over $30$
   random configurations of $\rho^{(m)}(\xt)$ for $N=700$ and $300$,
   and over $50$ configurations for $N=100$, $32$, $16$ and $8$.  The
   MV model parameter is chosen as $g^2\mu L=120$.\ \ Right:
   Chromo-electric and chromo-magnetic fields~\eqref{eq:E2} and
   \eqref{eq:B2} for the same MV model parameters and $N=700$.}
\label{fig:energy}
\end{figure}

\begin{figure}
 \begin{center}
 \includegraphics[width=0.49\textwidth]{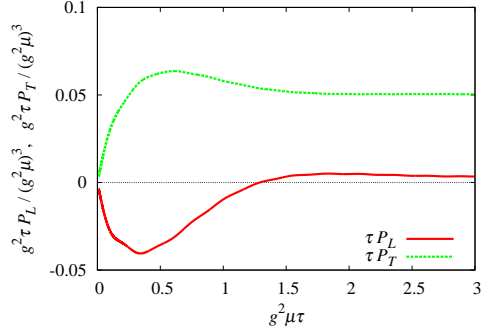}
 \end{center}
 \caption{Longitudinal pressure $P_{_L}$ defined in
   Eq.~\eqref{eq:P_{_L}} and transverse pressure $P_{_T}$ defined in
   Eq.~\eqref{eq:P_{_T}}, multiplied by $g^2\mu\tau$.}
\label{fig:P}
\end{figure}


Figure~\ref{fig:energy} shows the time evolution of the energy
density~\eqref{eq:E2} and \eqref{eq:B2}.  We set $g^2\mu L=120$ which
means that $g^2\mu R_A=67.7$ with $\pi R_A^2=L^2$ according to the
convention for the RHIC physics in the literature~\cite{Lappi:2003bi}.
This choice corresponds to $R_A\approx 7\fm$ [$\simeq
1.2\times(197)^{1/3}\fm$ (i.e. the radius of the Au atom)] and
$g^2\mu\approx 2\GeV$ (where $g=2$ as usually chosen, meaning that
$\alpha_s\simeq 0.3$).  In most of the numerical results presented in
this work, we keep using these parameters.  We note that the strength
of the Glasma instability as seen in the next section is set by the
instability seed $\Delta$ relative to $g^2\mu$ which characterizes the
CGC background.  In this sense, for our purpose in later discussions,
the precise value of $g^2\mu L$ is of no qualitative importance.

In Fig.~\ref{fig:energy}, we investigate the dependence on the
transverse grid size $N$:  The left plot shows how the energy density
increases with increasing $N$, varying from $N=8$ to $N=700$ from the
bottom curve to the topmost one.  It is already known in the
literature~\cite{Lappi:2003bi} that the $N$-dependence (or the
UV-cutoff dependence) is significant at $g^2\mu\tau\ll 1$, but becomes
almost irrelevant for $g^2\mu\tau\gtrsim 1$ (except for small lattice sizes
$N\ll 100$).

As is well-known from earlier
works~\cite{Krasnitz:1998ns,Krasnitz:1999wc}, the energy density
decreases as $\tau^{-1}$, which is the scaling behavior of a
free-streaming expansion.  In fact the energy density multiplied by
the time, $\tau\varepsilon$, saturates at a time $g^2\mu\tau\simeq 1$
(which is sometimes called the formation time, i.e. the time at which
partons are freed from the wave-function of the nuclei) where the
free-streaming expansion starts. This is natural on physical grounds,
because the classical fields keep interacting strongly until
$g^2\mu\tau\simeq 1$, and become quasi-free at $g^2\mu\tau\gg 1$.

We see that the longitudinal chromo-electric and chromo-magnetic
fields behave in the same way, which is the characteristic feature of
the Glasma initial state~\cite{Lappi:2006fp}.  In our convention
$E_{_L}^2$, $B_{_L}^2$, $E_{_T}^2$, and $B_{_T}^2$ are all approaching
a common value for $g^2\mu\tau>1$, and the system still remains
anisotropic (isotropy requires that $E_{_T}^2=2E_{_L}^2$ and
$B_{_T}^2=2B_{_L}^2$).  This becomes clearer if we consider the
transverse pressure $P_{_T}$ and the longitudinal one $P_{_L}$ defined
in Eqs.~\eqref{eq:P_{_T}} and \eqref{eq:P_{_L}}, respectively, as
shown in Fig.~\ref{fig:P}.  Then, we can find that $\tau P_{_T}$ and
$\tau P_{_L}$ both approach an asymptotic value and $\tau P_{_L}$ is
much smaller than $\tau P_{_T}$.  If $P_{_L}$ is exactly zero,
this means that the system reaches the limit of the free-streaming
expansion leading to $\varepsilon\propto \tau^{-1}$.  If the system is
completely thermalized, on the other hand, the pressure must be
isotropic and thus $P_{_T}=P_{_L}$.  Because an expansion with
positive $P_{_L}$ produces work against the expansion of the matter,
the energy density decreases faster than in the free-streaming case,
e.g. $\varepsilon\propto \tau^{-4/3}$ if $P_{_L}=\epsilon/3$ and if
the expansion is purely longitudinal (i.e. Bjorken expansion).
Therefore, the (leading order) Glasma state alone cannot reach
isotropization; a candidate for causing the longitudinal pressure to
grow is the instability due to the $\eta$-dependent fluctuations that
occur in higher order corrections.


\subsection{Spectral Decomposition}
\label{sec:spectral}

Let us now consider the time evolution of the spectral composition of
the energy content.  That is, the energy is written in terms of
Fourier components as
\begin{equation}
 \varepsilon_{_E} = \int\frac{\rmd^2 k_\perp}{(2\pi)^2}\,
  \varepsilon_{_E}(k_\perp) ,\qquad
 \varepsilon_{_B} = \int\frac{\rmd^2 k_\perp}{(2\pi)^2}\,
  \varepsilon_{_B}(k_\perp)
\label{eq:eE}
\end{equation}
in terms of the continuum variables.  If written in terms of the
lattice variables, the transverse momenta take discrete values
$k_i=2\pi n_i/L$ and the integration is replaced by a summation over
$n_i$.  The energy density in the mode $k_\perp$ is given by
\begin{align}
 \varepsilon_{_E}(k_\perp) &\equiv \bigl\langle \tr\bigl[
  E^{\eta a}(-\kt)E^{\eta a}(\kt) + \tau^{-2}\bigl(
  E^{ia}(-\kt)E^{ia}(\kt) \bigr) \bigr]\bigr\rangle ,
\label{eq:eE_k}\\
 \varepsilon_{_B}(k_\perp) &\equiv \bigl\langle \tr\bigl[
  B^{\eta a}(-\kt)B^{\eta a}(\kt) + \tau^{-2}\bigl(
  B^{ia}(-\kt)B^{ia}(\kt) \bigr) \bigr]\bigr\rangle\; ,
\label{eq:eB_k}
\end{align}
where $E^{\eta a}(\kt)$, $E^{i a}(\kt)$, $B^{\eta a}(\kt)$, and $B^{i
  a}(\kt)$ are the Fourier transforms of the chromo-electric and
chromo-magnetic fields.  The above expressions are given in terms of
the continuum variables, but it is non-trivial how to re-express them
in terms of the link variables:  Equation~\eqref{eq:eE_k} for the
chromo-electric component is not changed, but we cannot use
Eq.~\eqref{eq:eB_k} as it is\footnote{This is an issue only because we
  want to decompose the magnetic energy in Fourier modes.  If we stay
  in configuration space, the expressions for $B_{_L}^2$ and
  $B_{_T}^2$ in Eq.~\eqref{eq:B2} are perfectly fine to calculate
  the chromo-magnetic energy -- the problem is that they are not
  simply a sum of squares.}.

It is of course possible to introduce the chromo-magnetic field using
the relation~\eqref{eq:approxF} in such a way that
\begin{align}
 B_{_L}^a &= F_{xy}^a \approx
  \frac{2}{\rmi g}\tr \bigl[ t^a \bigl( 1-U_{xy} \bigr) \bigr]
  \notag\\
 B_{x}^a &= F_{\eta y}^a \approx
  \frac{2}{\rmi g a_\eta}\tr \bigl[ t^a \bigl( 1-U_{\eta y} \bigr)
  \bigr]\; ,\quad
 B_{y}^a = F_{\eta x}^a \approx
  \frac{2}{\rmi g a_\eta}\tr \bigl[ t^a \bigl( 1-U_{\eta x} \bigr)
  \bigr]\; .
\label{eq:approxB}
\end{align}
This approximation works for sufficiently large number of the
transverse sites, i.e.\ $N\gtrsim 100$.  Then, we in principle have
two options here:  we abandon the information on
$\varepsilon_{_B}(k_\perp)$ and just evaluate
$\varepsilon_{_E}(k_\perp)$ to see the spectral pattern, or, we adopt
a sufficiently large value of $N$ so that we can use the
approximation~\eqref{eq:approxB}.  In this work we take the former
option, because simulations with $N\gtrsim 100$ would be too
time-consuming in the forthcoming studies that include
$\eta$-dependent fluctuations (and where one thus needs to solve
($3+1$) dimensional equations).

At $\tau=0$, the initial spectrum is calculable analytically in the MV
model, leading to the expression~\cite{Fujii:2008km},
\begin{equation}
 \varepsilon = \frac{3(g^2\mu)^4}{2\pi g^2} \int
  \frac{\rmd^2 \kt}{(2\pi)^2} \frac{1}{k_\perp
  \sqrt{k_\perp^2+4m^2}}\ln\Biggl[ \frac{\sqrt{k_\perp^2+4m^2}+k_\perp}
  {\sqrt{k_\perp^2+4m^2}-k_\perp} \Biggr] ,
\label{eq:eE_pert}
\end{equation}
with an IR cutoff $m$. This formula means that
$\varepsilon_{_E}(k_\perp)$ behaves as $ k_\perp^{-2}\ln(k_\perp/m)$ at
large $k_\perp$, which is the expected perturbative tail.


\begin{figure}
 \begin{center}
 \includegraphics[width=0.49\textwidth]{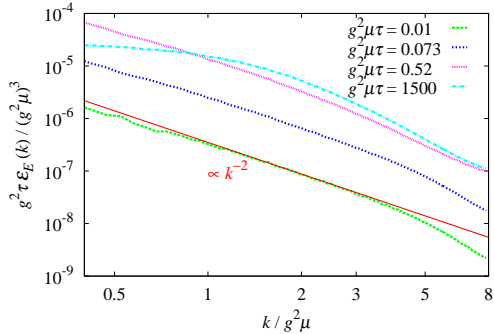}
 \end{center}
 \caption{Energy spectrum $\tau\varepsilon_{_E}(k_\perp)$ at
   $g^2\mu\tau=0.01$, $0.073$, $0.52$, and $1500$.  The transverse
   size is chosen as $N=300$ and the MV model parameter as $g^2\mu
   L=120$.  The spectrum at $g^2\mu\tau=0.01$ shows the perturbative
   scaling $\varepsilon_{_E}(k_\perp)\propto k_\perp^{-2}$.  An
   ensemble average is taken over $30$ configurations.}
 \label{fig:time-trans}
\end{figure}


At non-zero $\tau$, we have evaluated $\varepsilon_{_E}(k_\perp)$
numerically, with results shown in Fig.~\ref{fig:time-trans}.  To draw
the figure, we have calculated
$\varepsilon_{_E}(k_\perp=\sqrt{k_x^2+k_y^2})$ first as a function of
$k_x=2\pi n_x/L$ and $k_y=2\pi n_y/L$ and then have taken an average
over the results that lie in a bin, $N<\sqrt{n_x^2+n_y^2}<N+1$, to
obtain $\varepsilon_{_E}(k_\perp)$ with $k_\perp=2\pi N/L$.  We use
logarithmic horizontal and vertical axes in the plot of
Fig.~\ref{fig:time-trans}, to facilitate the identification of
the initial-time power-law scaling in the spectrum.

We can notice that the energy spectrum at the initial time
$g^2\mu\tau=0.01$ (where we started the simulation) obeys the
perturbative scaling $\sim k_\perp^{-2}$ up to the UV-cutoff, as
indicated by a line for eye guide in Fig.~\ref{fig:time-trans} -- in
agreement with the perturbative expression in Eq.~\eqref{eq:eE_pert}.
The spectral shape quickly approaches the asymptotic one within
$g^2\mu\tau\sim 1$ and it hardly changes after that time.  In fact the
shape at $g^2\mu\tau=0.52$ is already close to that at
$g^2\mu\tau=1500$ (where we stopped the simulation).  Thanks to this,
we do not show the results at any intermediate time between
$g^2\mu\tau=0.52$ and $g^2\mu\tau=1500$ in order to keep the plot
legible.

In later discussions in the section~\ref{sec:Lspectrum} we will
perform a similar analysis in the longitudinal direction to find
non-trivial asymptotic behavior of the spectral shapes.


\section{Breaking the Boost Invariance}
\label{sec:notBI}

After this summary of the boost-invariant results, including the time
evolution of the energy spectrum, let us proceed with the analysis of
the instability with respect to $\eta$-dependent fluctuations.  We
will first confirm the existence of the instability in a similar setup
as the one used in previous studies.  For this purpose, we choose the
initial fluctuations according to the prescription used in
Refs.~\cite{Romatschke:2005pm,Romatschke:2006nk} and we fix the MV
model parameter $g^2\mu$ in the same way as in the boost-invariant
situation.  We take the extent in the $\eta$ direction to be
$L_\eta=2$ units of rapidity and we use periodic boundary conditions
for $\eta$ as well as for transverse coordinates.  We note that this
extent covers roughly the mid-rapidity region where the
boost-invariant plateau is expected, and therefore it is legitimate to
have a boost-invariant background.  That is, our parameter choice
reads;
\begin{equation}
 g^2\mu a=\frac{120}{N}\quad (g=2)\;,
 \qquad  a_\eta = \frac{2.0}{N_\eta}\;,
\end{equation}
and we perform calculations at various values of the transverse grid
size $N$ and the longitudinal grid size $N_\eta$ later.

Introducing an arbitrary function $f(\eta)$, we can write the
$\eta$-dependent chromo-electric fields in such a way that Gauss's law
is satisfied by construction,
\begin{align}
 \delta E^i(x) &= a_\eta^{-1} \bigl[f(\eta-a_\eta)-f(\eta)\bigr]\,
  \xi^i(\xt) \;,
\label{eq:deltaEi}\\
 \delta E^\eta(x) &= -f(\eta)\sum_{i=x,y}
  \bigl[ U_i^\dagger(x-\hat{i})\xi^i(\xt-\hat{i})U_i(x-\hat{i})
  -\xi^i(\xt) \bigr]\;.
\label{eq:deltaEeta}
\end{align}
Before applying Eq.~\eqref{eq:deltaEeta} we can also add
$\eta$-dependent fluctuations in $U_i$.  For the moment we will
discuss the case with $\delta E^i$ and $\delta E^\eta$ only.  Here the
choice of $\xi^i(\xt)$ is arbitrary and we choose it as a random
variable in the transverse plane, i.e.\
\begin{equation}
 \langle \xi^i(\xt)\xi^j(\xt')\rangle
  = \delta^{ij}\delta^{(2)}(\xt-\xt') \;.
\end{equation}
It is important to note that $\xi^i$ has the dimension of a momentum,
which is obvious from the above equation.  If all the dimensional
quantities are scaled with the transverse spacing $a$ as is the case
in the numerical implementation, $\xi^i$ scales as $a^{-1}$, which is
artificial.  The strength of the longitudinal disturbance should be
given independently.  To cancel this artificial $a$ dependence, 
we use a dimensionless $\bar{\xi}^i$ satisfying
$\langle\bar{\xi}^i(\xt)\bar{\xi}^j(\xt')\rangle
=\delta^{ij}\delta_{x,x'}$ (on the lattice) and make $f(\eta)$ scale
as $N^{-1}$, so that the fluctuations are proportional overall to
$(Na)^{-1}=L^{-1}$ which is a constant.

In the literature~\cite{Romatschke:2005pm,Romatschke:2006nk} $f(\eta)$
is chosen as a random variable too containing all infrared and
ultraviolet modes in the spectrum.  Here, in order to study the
instability behavior in a well-controlled situation, let us
perturb the system with a single $\eta$-mode, and consider
superposition of multiple modes only later\footnote{It is legitimate
  to do so as long as the perturbations are small compared to the
  background field, so that their dynamics remains linear. After that,
  the different fluctuation modes will evolve non-linearly and mix.
  We will return to this point later.}.
Hence, we will adopt the following simple form for the
moment\footnote{In this lattice parameterization of $f(\eta)$, the
  frequency $\nu_0$ is an integer between $-N_\eta/2$ and
  $+N_\eta/2$.},
\begin{equation}
 f(\eta) = \Delta \cos\Bigl( \frac{2\pi\nu_0}{L_\eta}\eta \Bigr)\;,
\label{eq:feta}
\end{equation}
where $L_\eta=N_\eta a_\eta$ is chosen to be $2$ as we explained
before and $\Delta$ should scale as
\begin{equation}
 \Delta = \frac{\Delta_0}{N} \;,
\label{eq:Delta0}
\end{equation}
to make the fluctuations insensitive to the way we discretize the
transverse plane.  We will later fix $\Delta_0$ and then vary $N$
to study the (unphysical) dependence on the transverse grid size.


\subsection{Dependence on the transverse grid size $N$}

Here we fix $\nu_0=1$ in Eq.~\eqref{eq:feta} because, as we will
confirm later, this lowest non-zero mode leads to the fastest and
strongest growth in the longitudinal pressure.

\begin{figure}
 \includegraphics[width=0.49\textwidth]{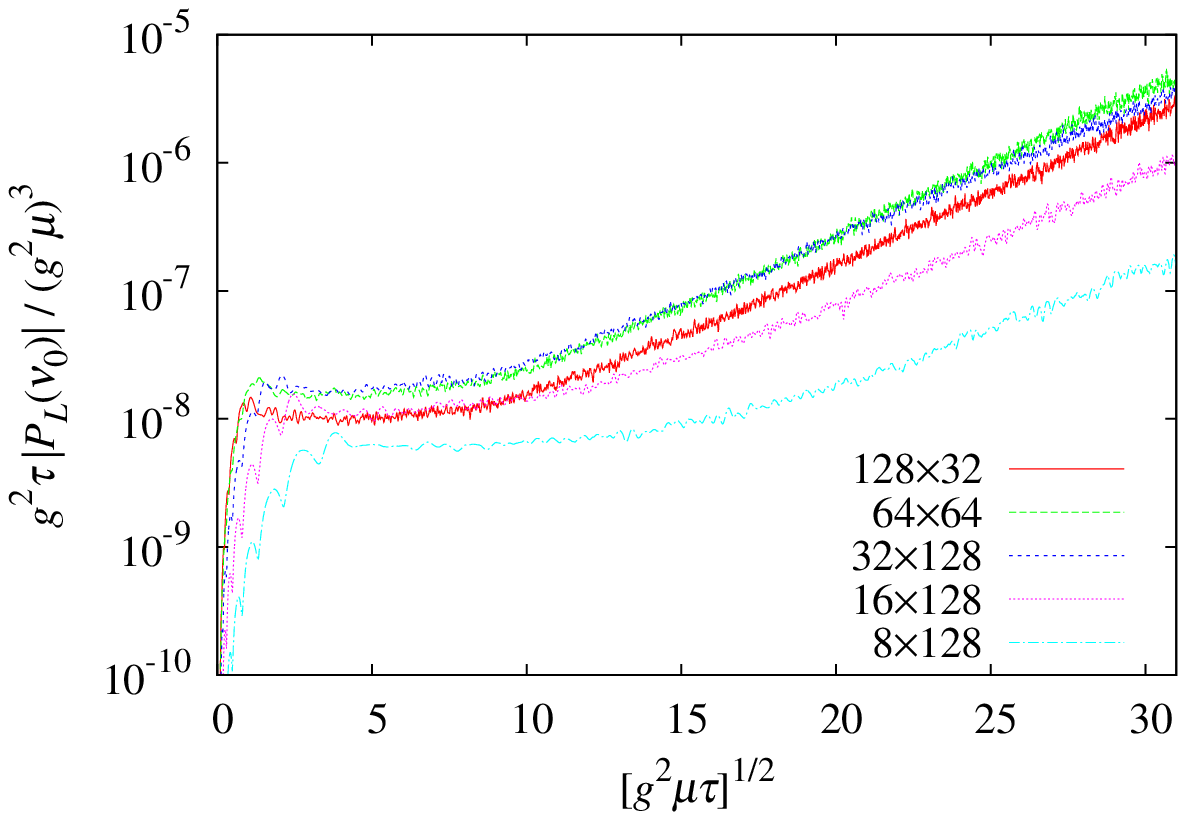}
 \includegraphics[width=0.49\textwidth]{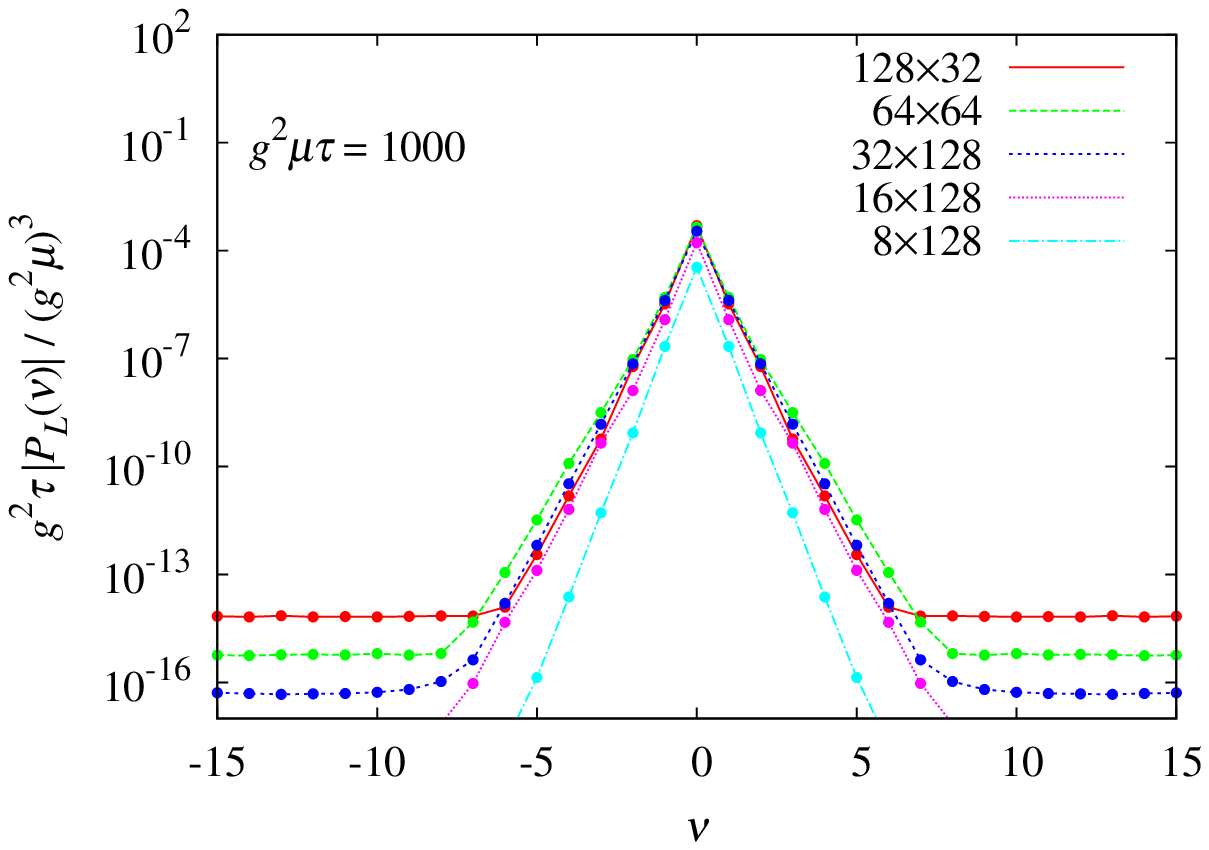}
 \caption{Left: Longitudinal ``pressure'' at $\nu=\nu_0$, where
   $\nu_0=1$ is chosen for the initial condition~\eqref{eq:feta}.  The
   results are for $N=8$, $16$, $32$, $64$, $128$ in the transverse
   direction (from the top to the bottom) and $N_\eta=128$ fixed in
   the longitudinal direction (except for $N_\eta=64$ for $N=64$ and
   $N_\eta=32$ for $N=128$).  The seed magnitude is $\Delta=10^{-5}$
   for $N=32$, i.e.\ $\Delta_0=32\times 10^{-5}$ in
   Eq.~\eqref{eq:Delta0}.  An ensemble average is taken over $100$
   configurations.\ \ Right: Spectrum of the longitudinal
   ``pressure'' as a function of $\nu$ at
   $g^2\mu\tau=1000$ ($\sqrt{g^2\mu\tau}\approx 32$).} 
 \label{fig:size_t}
\end{figure}

Figure~\ref{fig:size_t} shows how the instability occurs in the
longitudinal pressure $P_{_L}$ in the long-time run.  The left-panel is
the pressure multiplied by time ($g^2\tau P_{_L}$) in the unit of
$(g^2\mu)^3$.  We note that the plotted quantity is for $\nu=\nu_0=1$
which maximizes the Fourier transform of $\tau P_{_L}$ (except for the
zero mode $\nu=0$).  That is, we define,
\begin{equation}
 P_{_L}(\nu) \equiv \frac{1}{L^2}\int\rmd^2 \xt\,
  \frac{1}{L_\eta}\int_0^{L_\eta}\rmd\eta\,
  P_{_L}(\eta,\xt)\, \rme^{\rmi (2\pi\nu/L_\eta)\eta} \;,
\end{equation}
which is complex valued in general\footnote{We note that $\nu$ in our
  definition ranges from $-N_\eta/2$ to $N_\eta/2$ and the
  corresponding wave-number is $k_\eta = 2\pi\nu/L_\eta$.}.  We thus
plot its modulus, $|g^2\tau P_{_L}(\nu=\nu_0)/(g^2\mu)^3|$, in the
left panel of Fig.~\ref{fig:size_t} corresponding to an initial
disturbance at $\nu_0=1$ and $\Delta_0=32\times 10^{-5}$ in
Eqs.~\eqref{eq:feta} and \eqref{eq:Delta0} (i.e.\ we choose $N=32$ as
a reference point and then $\Delta=10^{-5}$ for $N=32$).  The
horizontal axis represents $\sqrt{g^2\mu\tau}$, which is a natural
variable in a longitudinally expanding geometry.  Note that the
Fourier transform $P_{_L}(\nu)$ at finite $\nu$ is not a measurable
quantity in experiments.  Its real and imaginary parts both fluctuate
from negative to positive values depending on the initial conditions
and they would be vanishing if we compute an average without taking
the modulus.  Only the $\nu=0$ mode has a physical meaning as the
longitudinal pressure.  For our purpose of studying the instability, we
can nevertheless consider $\left<|P_{_L}(\nu)|\right>$ (i.e. we
compute the modulus first, and we perform the average over initial
conditions next) in order to confirm some known results reported
previously~\cite{Romatschke:2005pm,Romatschke:2006nk}.  In later
discussions, we will also decompose the energy density into $\nu$
components, which is a more well-defined quantity to see the
instability.


From the left panel of Fig.~\ref{fig:size_t}, the exponential growth
with the square root of proper time is obvious.  The onset of the
instability depends on the transverse size $N$ but the instability
slope is rather insensitive to $N$.  With increasing $N$, the
instability occurs earlier, and eventually the shift in this onset
position saturates for $N\approx 32$.  This means that the instability
does not start earlier by increasing $N$ further.  So, we stop
increasing $N$ and conclude that simulations at $N=32$ are reasonably
close to the continuum limit. In the right panel, we see how the
fields, initially localized in the modes $\nu_0=0$ (background field
itself) and $\nu_0=\pm 1$ (perturbation), populate the higher $\nu$
modes due to the non-linearities in the Yang-Mills equations.  For a
small perturbation, the amplitude in the higher modes decreases very
fast (exponentially with $\nu$), because in order to produce a field
at a given $\nu$, one needs to multiply $\nu$ seeds at $\nu_0=1$.


\subsection{Dependence on the longitudinal grid size $N_\eta$}


\begin{figure}
 \includegraphics[width=0.49\textwidth]{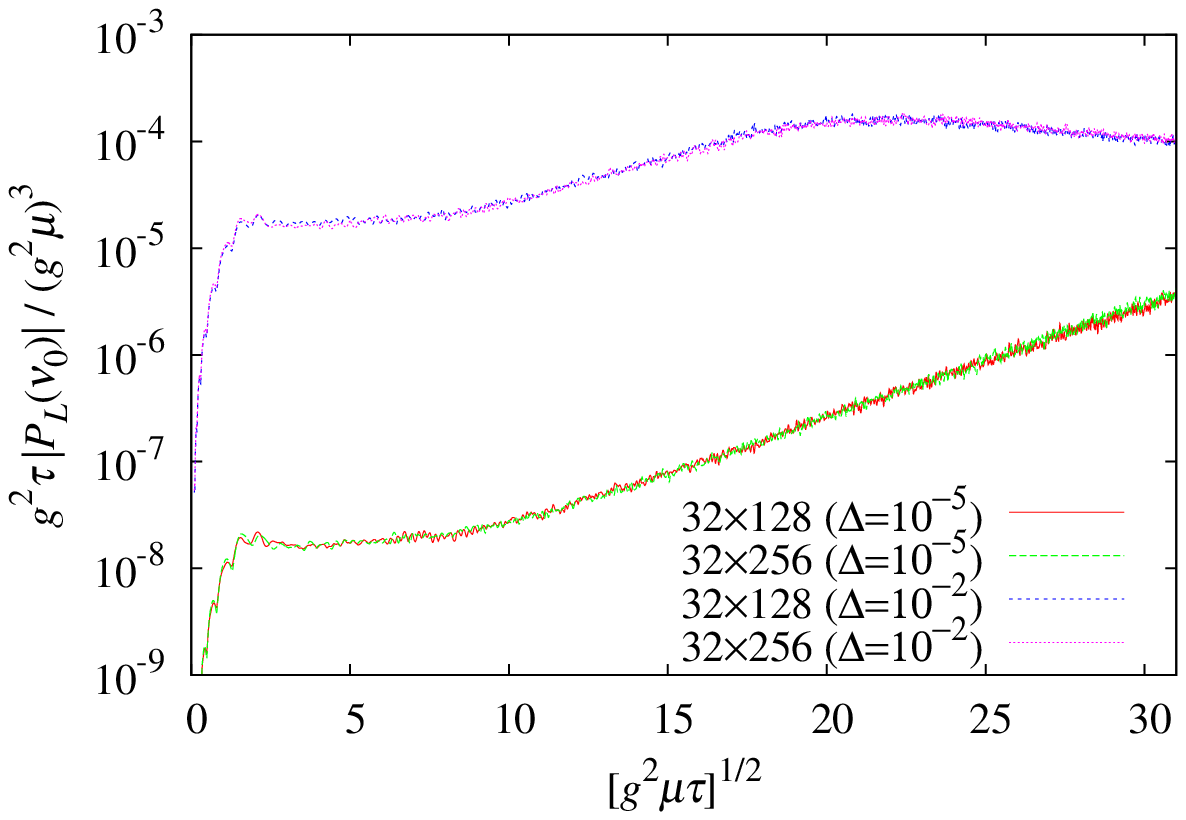}
 \includegraphics[width=0.49\textwidth]{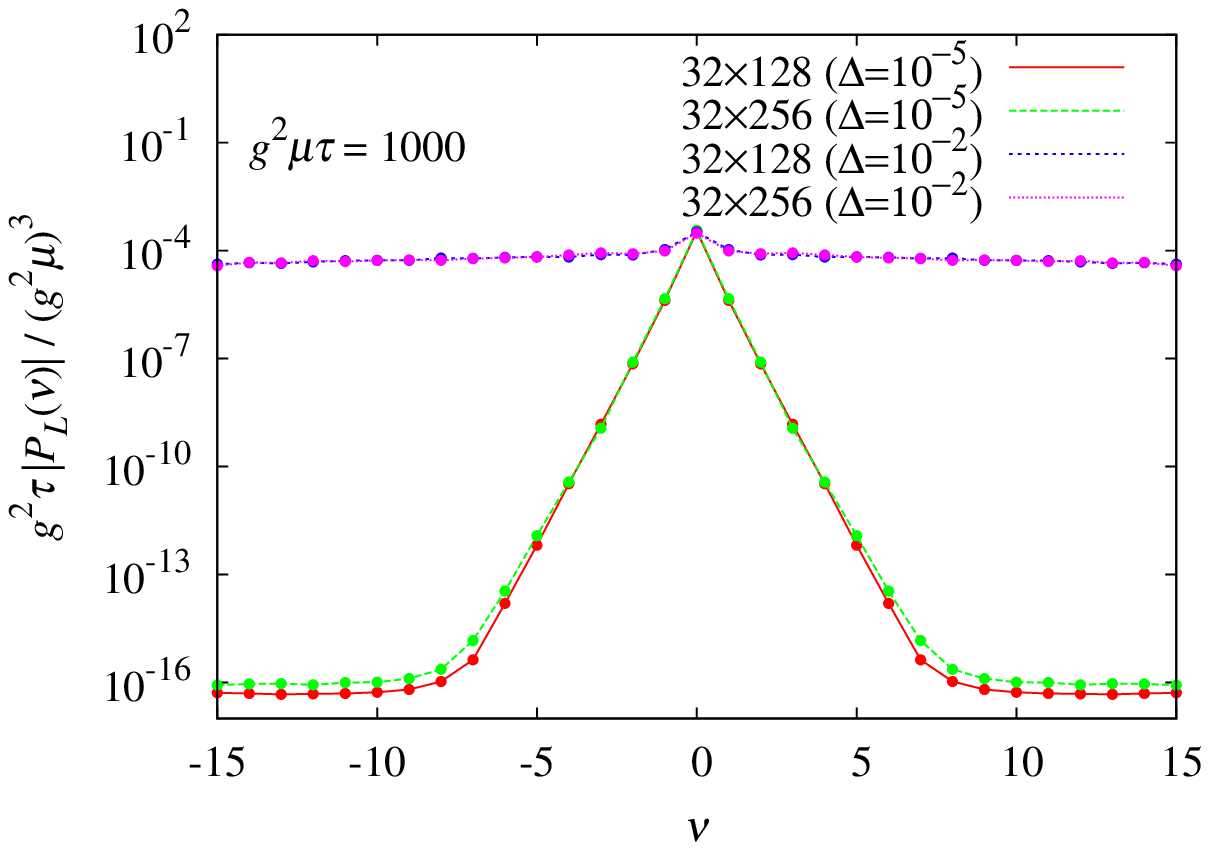}
 \caption{Left: Longitudinal ``pressure'' at $\nu=\nu_0=1$.  The
   results are for $\Delta=10^{-5}$ and $N=32$ fixed in the transverse
   direction and $N_\eta=128$ and $N_\eta=256$ in the longitudinal
   direction.  An ensemble average is taken over $100$
   configurations.\ \ Right: Spectrum of the longitudinal
   ``pressure'' as a function of $\nu$ at
   $g^2\mu\tau=1000$.}
 \label{fig:size_l}
\end{figure}


Physical results must also not depend on how we discretize the
longitudinal coordinate.  In order to see the dependence on the
longitudinal grid size, we make a plot in Fig.~\ref{fig:size_l} in the
same way as in Fig.~\ref{fig:size_t}, where we compare two
longitudinal grid sizes that differ by a factor $2$.  As is clear from
Fig.~\ref{fig:size_l} we cannot find any sizable dependence on
$N_\eta$, which means that our numerical calculations are safely free
from lattice artifacts in the longitudinal direction.

The influence of $N_\eta$ can be seen more clearly in the spectrum of
$|P_{_L}(\nu)|$ shown in the right panel of Fig.~\ref{fig:size_l}.  A
different choice of $N_\eta$ with fixed $L_\eta=N_\eta a_\eta$ changes
the UV cutoff (the spectrum ranges between $-N_\eta/2$ and
$+N_\eta/2$): in the right panel of Fig.~\ref{fig:size_l}, $\nu$
varies from $-64$ to $+64$ for the $N_\eta=128$ calculation and from
$-128$ to $+128$ for the $N_\eta=256$ calculation.  Since the seeds
are at $\nu=\nu_0=1$ in our initial condition, we naturally anticipate
that the UV cutoff at $\nu=N_\eta/2$ does not affect physical
properties around $\nu\sim\nu_0\ll N_\eta/2$.

One may think that the results for a larger seed amplitude
$\Delta=10^{-2}$ in Fig.~\ref{fig:size_l} could be sensitive to the
choice of $N_\eta$ because all the amplitudes in the spectrum are
substantially larger and thus a change in $N_\eta$ leading to a change
in the UV cutoff may have an influence on the spectrum.  We have also
checked this and found that it is not the case.  As is clear from
Fig.~\ref{fig:size_l}, the difference between the results for
$32\times 128$ with $\Delta=10^{-2}$ and that for $32\times 256$ with
$\Delta=10^{-2}$ is almost invisible.

From all these results we can conclude that a value of $N_\eta=128$ is
sufficiently large as long as $\nu_0$ is set to be of order unity.  If
$\nu_0$ takes a larger value, as we will see soon below, the spectral
shape has a wider distribution and the UV cutoff at $\nu=N_\eta/2$
could influence the results.


\subsection{Dependence on the magnitude $\Delta$ of the seed}


\begin{figure}
 \includegraphics[width=0.49\textwidth]{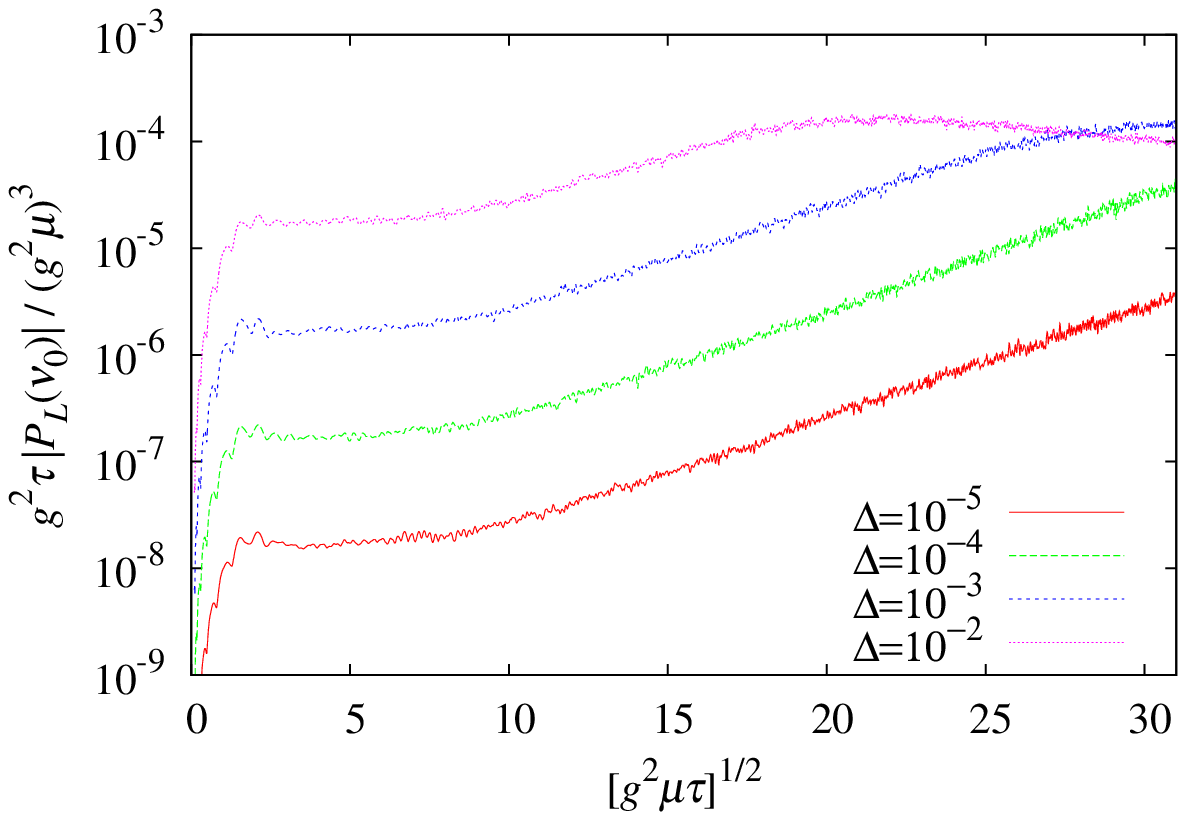}
 \includegraphics[width=0.49\textwidth]{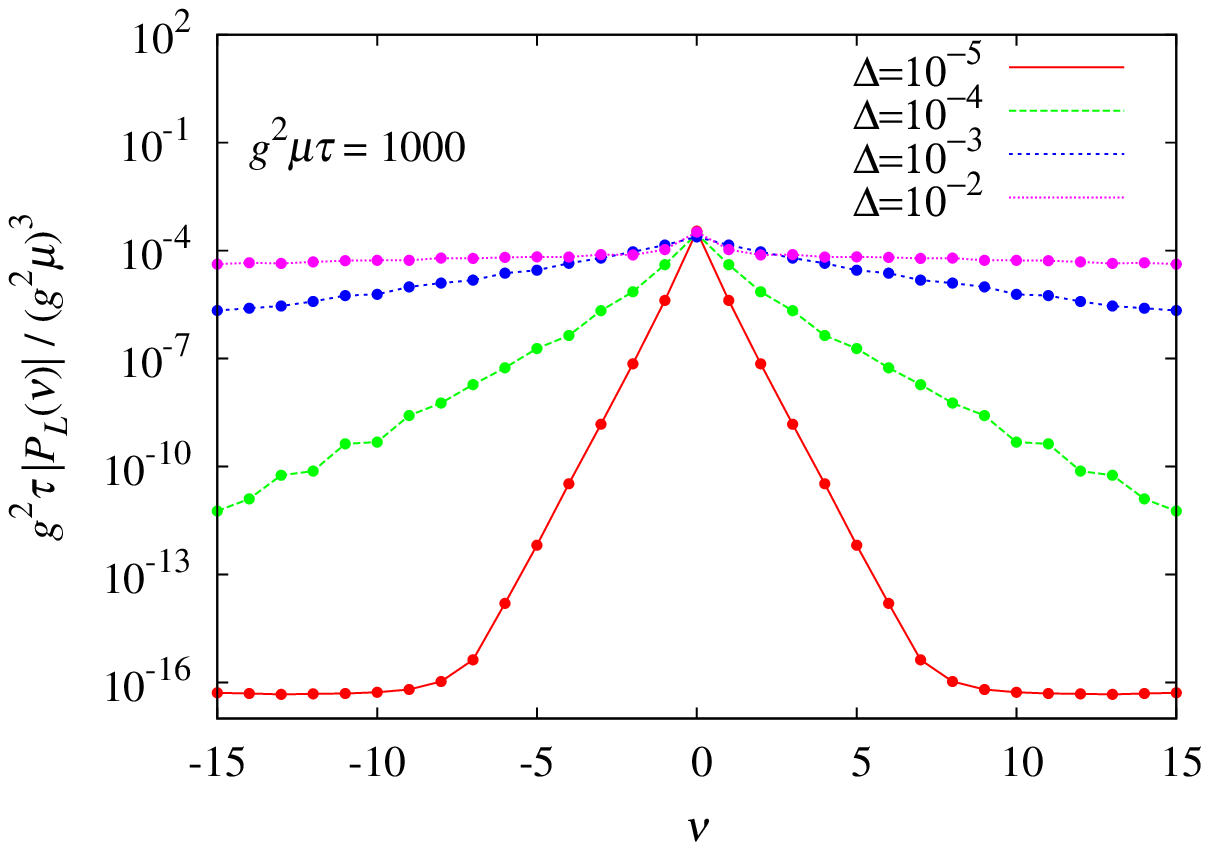}
 \caption{Left: Longitudinal ``pressure'' at $\nu=\nu_0=1$.  The
   results are for $\Delta=10^{-3}$, $10^{-4}$, $10^{-5}$ and a fixed
   lattice size $32\times 128$.\ \
   Right: Spectrum of the longitudinal ``pressure'' as a function of
   $\nu$ at $g^2\mu\tau=1000$, which has a UV cutoff at
   $\nu_{\text{max}}=64$ for the $N_\eta=128$ calculation.}
 \label{fig:magnitude}
\end{figure}


In this subsection, we study the dependence on the seed magnitude
$\Delta$ of the longitudinal fluctuations in Eq.~\eqref{eq:feta}.  In
Fig.~\ref{fig:magnitude}, we show some of numerical results for
different values of $\Delta$.  It is interesting that the results
obviously scale as $|P_{_L}|\propto \Delta$ up to the point where the
instability growth eventually saturates.  This $\Delta$ dependence is
almost trivial, which suggests that the non-linear effects in terms of
finite-$\nu$ modes are small unless the unstable modes substantially
develop at large $g^2\mu\tau$.  One might have thought that any
physical observables should be expanded in even powers of $\Delta$ and
thus $|P_{_L}(\nu_0)|$ should be proportional $\Delta^2$ rather than
$\Delta$.  This is not the case because we took the modulus of complex
$P_{_L}(\nu_0)$ before performing the average over initial conditions,
thus preventing the linear term in $\Delta$ from vanishing.

We can also notice from the right panel of Fig.~\ref{fig:magnitude}
that, once the instability at $\nu=\nu_0$ gets saturated, the mode at
$\nu=\nu_0$ stops growing but the spectrum spreads quickly to higher
$\nu$-modes.  Once this happens, the scaling property with
  $\Delta$ is lost.  Actually the amplitude of the $\nu=\nu_0$ mode
reaches one third of the zero-mode amplitude (for $\Delta=10^{-2}$
results at $g^2\mu\tau=1000$) before this saturation occurs.  Then the
non-linearities or self-interaction effects cannot be neglected any
longer.


\subsection{Dependence on the initial seed wave-number $\nu_0$}

Now we investigate the $\nu_0$-dependence of the instability behavior.
First, we have chosen three different $\nu_0$'s as $\nu_0=1$, $5$, and
$10$ to uncover the general trend.  Then we study a more general case,
that consists in a superposition of these initial wave-numbers.


\begin{figure}[htb]
 \includegraphics[width=0.49\textwidth]{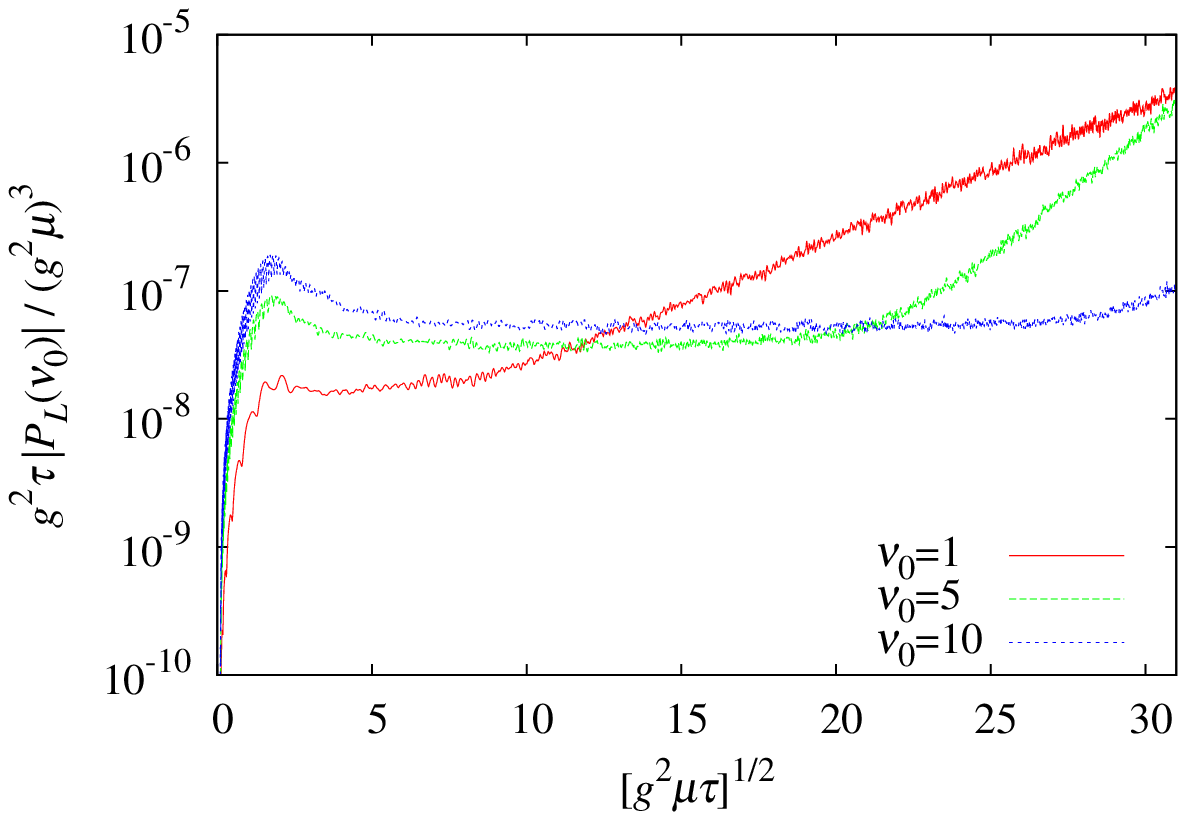}
 \includegraphics[width=0.49\textwidth]{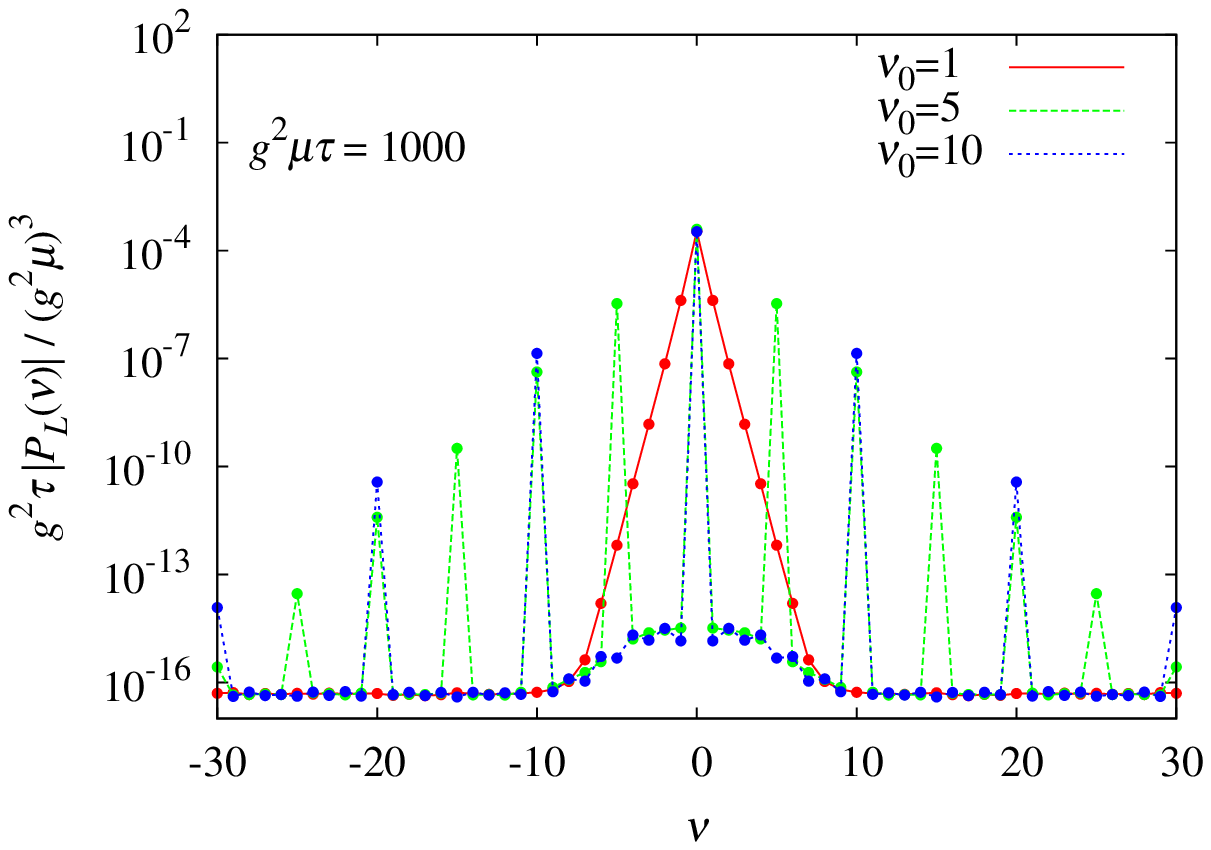}
 \caption{Left: Longitudinal ``pressure'' at $\nu=\nu_0$ with
   $\nu_0=1$, $5$, and $10$ in the initial condition~\eqref{eq:feta}.
   The lattice size $32\times 128$ is fixed and the seed amplitude is
   $\Delta=10^{-5}$.\ \ Right: Spectrum of the longitudinal
   ``pressure'' as a function of $\nu$ at $g^2\mu\tau=1000$.}
 \label{fig:nu}
 \includegraphics[width=0.49\textwidth]{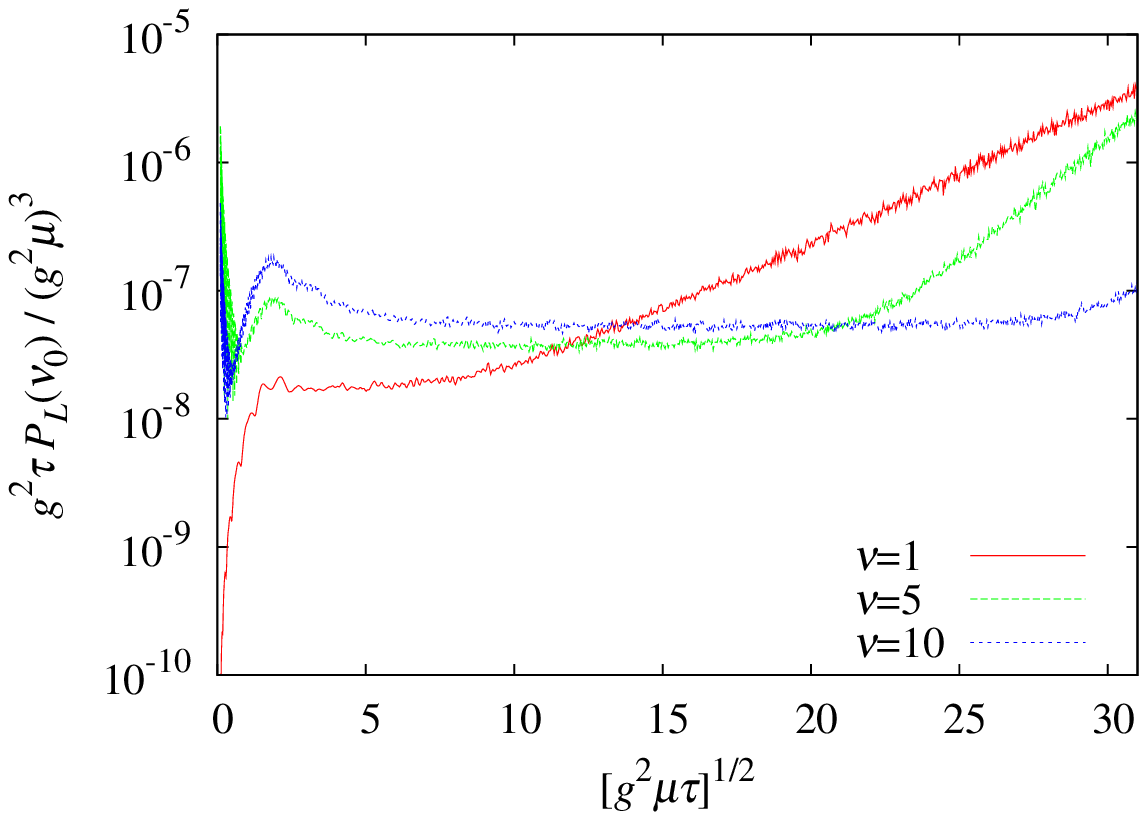}
 \includegraphics[width=0.49\textwidth]{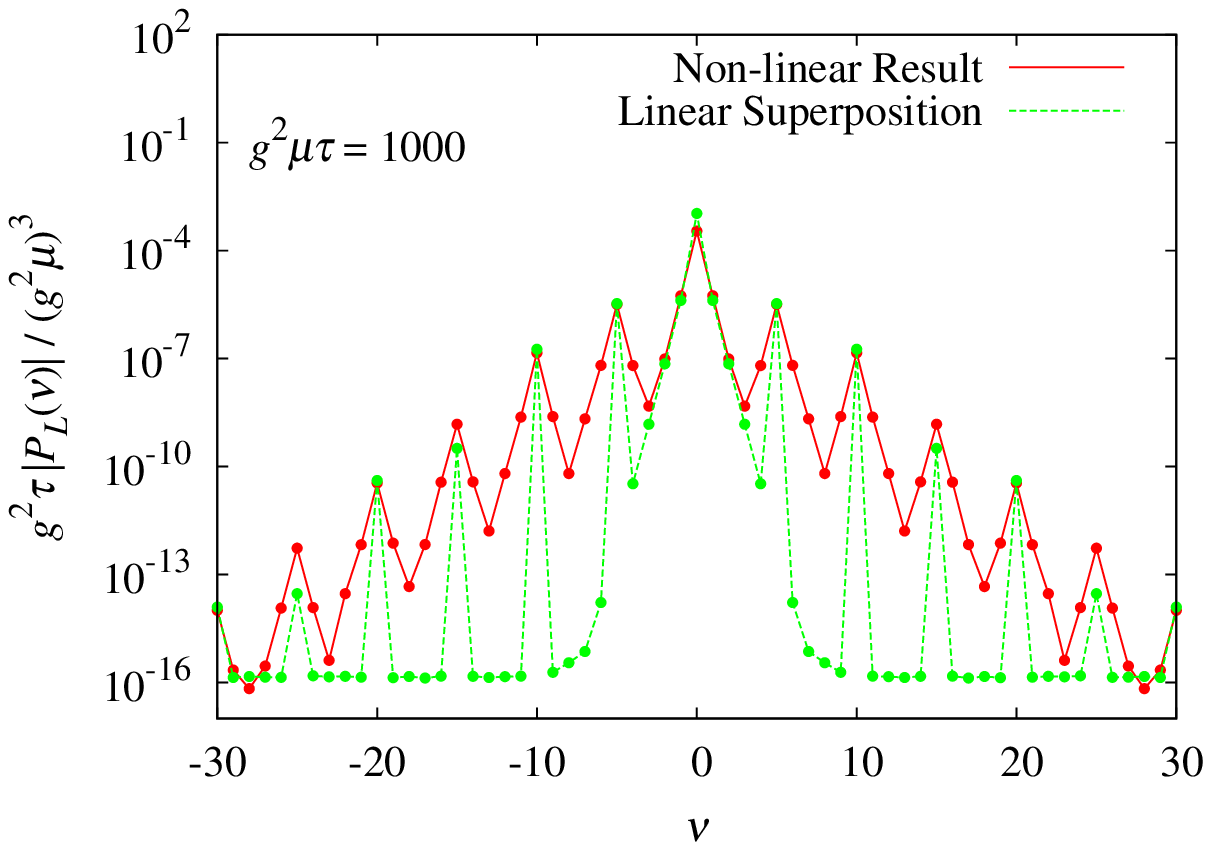}
 \caption{Left: Longitudinal ``pressure'' at $\nu=1$, $5$, and
   $10$ with the initial condition given as a
   superposition~\eqref{eq:feta_mix}.  Right: Spectrum of the
   longitudinal ``pressure'' as a function of $\nu$ at
   $g^2\mu\tau=1000$.  The solid line represents the results from
     the non-linear evolution with the initial condition given as a
     superposition, while the dotted one is a simple linear
     superposition of the results shown in Fig.~\ref{fig:nu}.}
 \label{fig:nu_mix}
\end{figure}


The figure~\ref{fig:nu} shows the numerical results.  It is clear from
the figure that the onset of the instability is delayed for larger
$\nu_0$.  This is naturally understandable from the kinetic term for
finite-$\nu$ modes.  The (positive) kinetic term generally tends to
stabilize the system, and so the (negative) term responsible for the
instability is overwhelmed if $\nu$ is large enough.  For dimensional
reasons, however, this kinetic term for the finite-$\nu$ modes is of
the form $\nu^2/\tau^2$, and therefore it decreases with time.  In
contrast, the negative term that generates the instability has a
constant coefficient of order unity -- therefore, it dominates over
the kinetic term at sufficiently large time. In other words, the
kinetic term in $\nu^2/\tau^2$ only delays the start of the
instability by a time proportional to $\nu$.  We can confirm
qualitatively this interpretation from Fig.~\ref{fig:nu}.  The growth
appears to start around $\sqrt{g^2\mu\tau}\sim 9.5, 21, 30$ for
$\nu_0=1, 5, 10$, respectively, and the starting times are indeed
ordered proportionally to $\nu_0$.

The spectral shape shown in the right panel of Fig.~\ref{fig:nu} looks
discontinuous for $\nu_0=5$ and $10$, but this is a trivial artifact
of the fact that our initial seed contains a single $\nu_0$ mode.
Indeed, the non-linear couplings in the Yang-Mills equations can only
produce linear combinations with integer coefficients of the
frequencies present in the initial condition.  For instance, with
$\nu_0=5$, only modes with $\nu$ a multiple of $5$ can be produced by
the non-linear evolution. The fact that the amplitude if these higher
harmonics is suppressed compared to the amplitude of the base
frequency $\nu_0$ is a good indication that we are still in a regime
where the non-linearities are rather small.

We can also see that the non-linear effects are certainly present but
not very large by repeating the above calculation with an initial
condition given as a superposition of three modes, namely,
\begin{equation}
 f(\eta) = \Delta \sum_i
  \cos\Bigl( \frac{2\pi\nu_0^{(i)}}{L_\eta}\eta \Bigr)
\label{eq:feta_mix}
\end{equation}
with $\nu_0^{(1)}=1$, $\nu_0^{(2)}=5$, $\nu_0^{(3)}=10$, instead of
the single-mode fluctuation of Eq.~\eqref{eq:feta} with $\nu_0=1$,
$\nu_0=5$, $\nu_0=10$ individually.  If the non-linearity is
significant, the results should not be a simple superposition of the
individual calculation.  In view of Fig.~\ref{fig:nu_mix}, however, we
can see only small deviations from the individual results
shown in Fig.~\ref{fig:nu} at $g^2\mu\tau=1000$.  This is an important
observation because the initial fluctuations in reality should have
some continuous spectral distribution (see
Ref.~\cite{Fukushima:2006ax} for instance) and we simplified our
analysis by considering the special case of a single-mode fluctuation.
As long as the non-linearity is a minor effect, we can just add up the
single-mode outputs in order to get approximate results for general
initial spectra.

Although they are several orders of magnitude smaller at
$g^2\mu\tau=1000$, we see that higher harmonics mixing the three
$\nu_0$'s are enhanced by non-linear effects in the right panel of
Fig.~\ref{fig:nu_mix}.  For instance, the amplitude of the mode
$\nu=21$ is much larger than the sum of the amplitudes at $\nu=21$ for
the three seeds considered in isolation.  Indeed, when the seed
$\nu_0=1$ is alone for example, the mode $\nu=21$ is produced only
after 20 interactions, which explains why it is totally negligible.
Furthermore, the seeds $\nu_0=5$ or $10$ can never lead to $\nu=21$ by
themselves.  When the three seeds are superposed in the initial
condition, 21 can be reached more efficiently as $10+10+ 1$, i.e. with
2 interactions only.

Apart from small non-linear effects, it is certainly true that the
smallness of the field fluctuations we have considered gives a
justification for our simplified, single mode, analysis.  However, at
the same time, thermalization is hardly expected to occur in these
circumstances, precisely because in this regime the interaction
effects are small.  In realistic conditions, the parameter $\Delta$
that controls the fluctuations is related to the strong coupling
$\alpha_s$, and therefore it could be larger than $\Delta_0=10^{-5}$
that we have considered here.  As a consequence, we expect that the
non-linearities would start much sooner and be much more important in
applications of this framework to actual heavy-ion collisions.  The
study with such large $\Delta$ is beyond our current scope because a
large $\Delta$ requires full information on quantum fluctuations and
needs appropriate renormalization procedures, which is technically
involved~\cite{Dusling:2011rz}.


\section{Time Evolution of the Longitudinal Spectrum}
\label{sec:Lspectrum}

In spite of the instability the zero-mode is still dominant among all
the modes in $P_{_L}(\nu)$.  With an initial condition that contains a
single mode as in Eq.~\eqref{eq:feta}, the $\nu=\nu_0=1$ mode is the
second largest, which is obvious from the spectrum shown in the right
panel of Fig.~\ref{fig:size_t}.  Moreover, the spectrum is spreading
into higher wave-numbers as the time elapses.  In this section,
we shall discuss the scaling property of the energy spectrum with
respect to the frequency $\nu$, and show that it possibly displays
also a Kolmogorov's turbulent spectrum.  A turbulent spectrum has
already been reported in the context of the QCD plasma
instability~\cite{Arnold:2005ef,Ipp:2010uy}, but these results
with the power $\simeq -2$ instead of $-5/3$ are in conflict with
Kolmogorov's wave turbulence.  On the other hand, the study performed
in Ref.~\cite{Berges:2008mr} has shown that a Kolmogorov spectrum
appears due to instabilities in Yang-Mills theory, for a fixed-volume
system.  Therefore, it is still worth addressing the turbulent
spectrum in the present Glasma simulation, that has longitudinal
expansion.  This is because the Glasma system is expanding along the
beam axis, which tends to tame the turbulences and to delay the
isotropization or thermalization in general.  It is thus non-trivial
whether the universal scaling still holds or not in the expanding
system.  Besides, our description of the Glasma evolution is purely
classical and the precise relationship between the Glasma instability
and the QCD plasma instability is not fully understood yet.


\begin{figure}
 \includegraphics[width=0.49\textwidth]{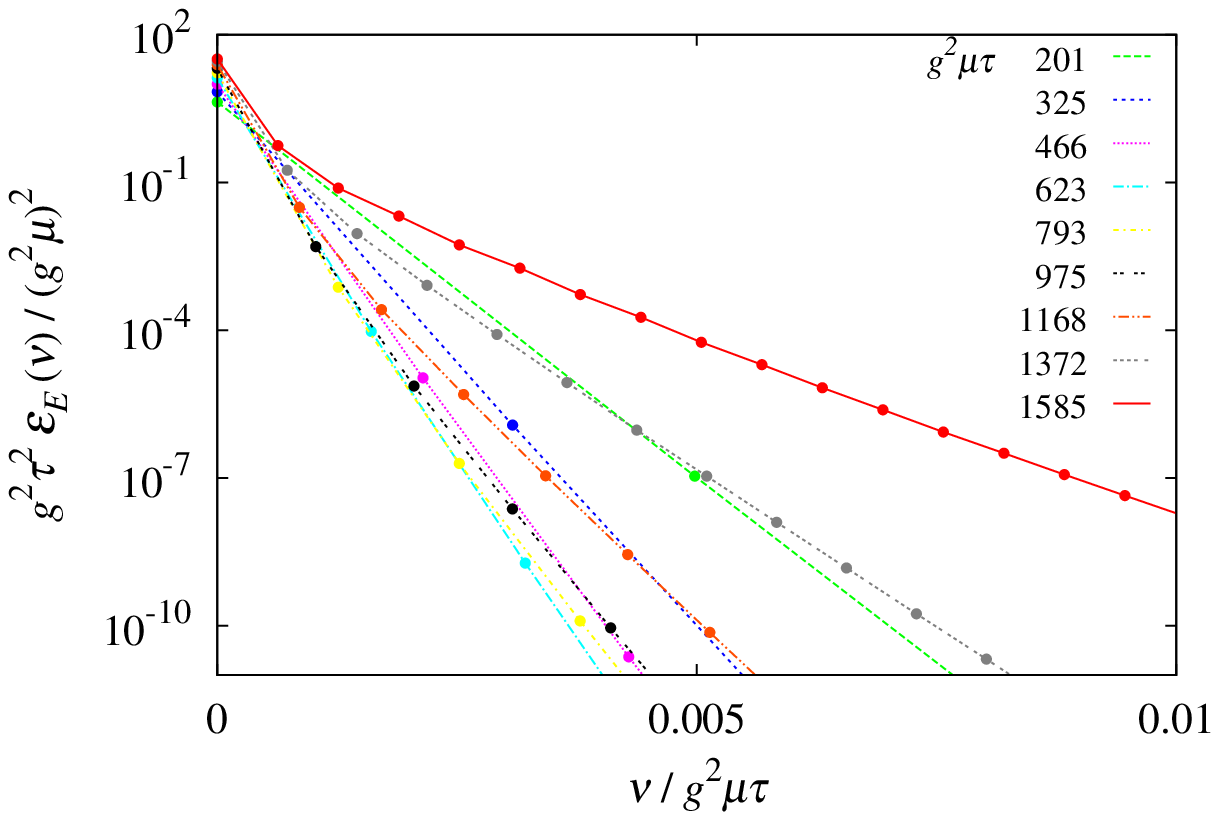}
 \includegraphics[width=0.49\textwidth]{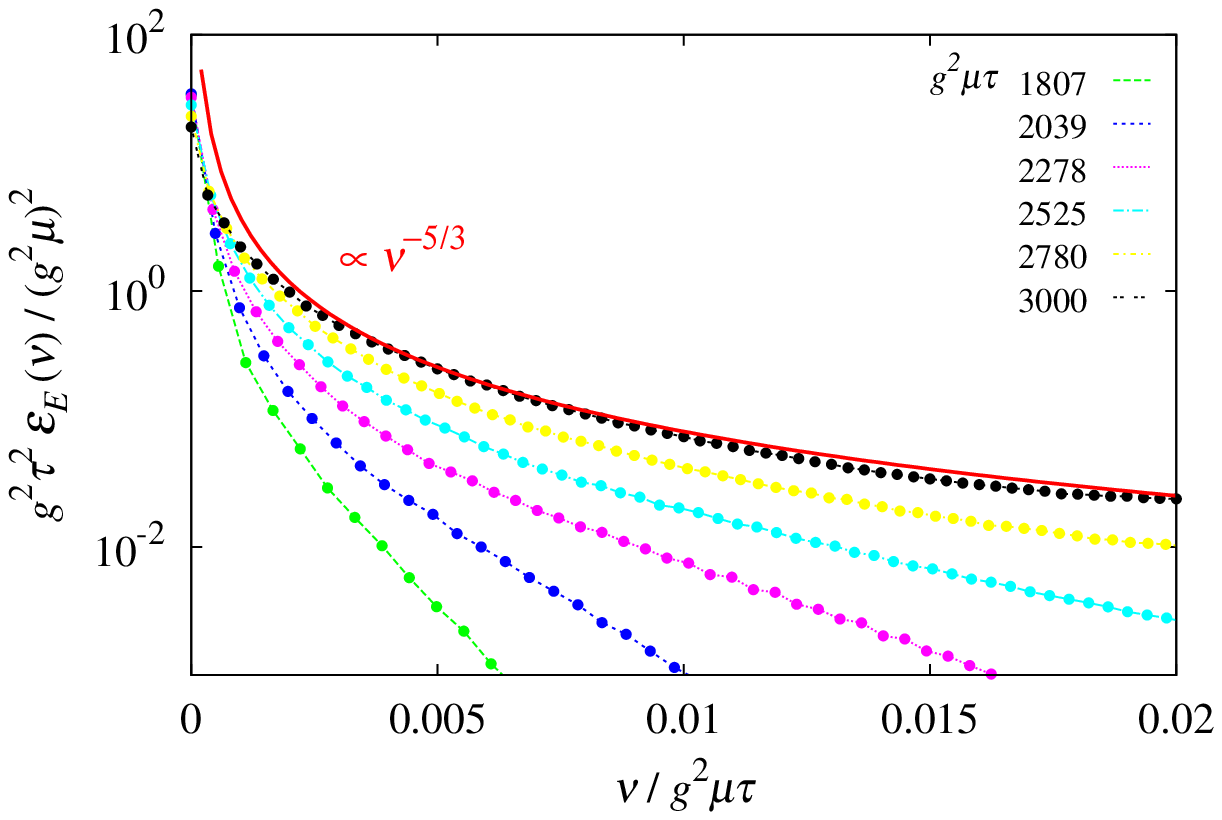}
 \caption{Time evolution of the chromo-electric energy density as
   a function of the scaled longitudinal wave-number
   $\nu/(g^2\mu\tau)$.  The lattice size is $32\times 128$ and the
   seed is $\Delta=10^{-5}$.\ \ Left: Results at early times up to
   $g^2\mu\tau \sim 1500$.  Note that the horizontal axis is not
   logarithmic and also that the vertical axis is for not
   $\tau\varepsilon_{_E}(\nu)$ but $\tau^2\varepsilon_{_E}(\nu)$ (see
   the text for details). \ \ Right: Results at later times up to
   $g^2\mu\tau=3000$.}
 \label{fig:time-scale-long}
\end{figure}


Two plots in Fig.~\ref{fig:time-scale-long} show the longitudinal
energy spectrum (for the chromo-electric part only), that is, the
energy density decomposed in terms of the longitudinal wave-number
$\nu$ as
\begin{equation}
 \varepsilon_{_E} = \int\frac{\rmd\nu}{2\pi} \varepsilon_{_E}(\nu)
\label{eq:eEnu}
\end{equation}
with
\begin{equation}
 \varepsilon_{_E}(\nu) \equiv \bigl\langle\tr\bigl[ E^{\eta a}(-\nu)
  E^{\eta a}(\nu) + \tau^{-2}\bigl( E^{ia}(-\nu)E^{ia}(\nu) \bigr)
  \bigr]\bigr\rangle .
\end{equation}
(The transverse momentum $\kt$ of the fields is integrated over.)  The
question is then what kind of scaling spectrum we can expect for
$\varepsilon_{_E}(\nu)$ at later times.  For this purpose, we consider
three characteristic quantities; the wave-number, the Fourier
decomposed energy, and the rate of energy flow in $\nu$-space
(denoted by $\psi$ below).  The wave-number $\nu$ itself is, however,
dimensionless and it seems difficult to cope with the dimensional
matching argument by Kolmogorov.  We propose that one should use a
scaled variable $\nu/(c\tau)$ instead of $\nu$, in the case of an
expanding geometry.  This is because $c\tau\eta$ is approximately
equal to the longitudinal coordinate $z$ if $\eta$ is small enough,
and $\nu/(c\tau)$ is approximately equal to the momentum $p_z$.  Then,
the dimension of the scaled wave-number is $[\nu/c\tau]=l^{-1}$.  We
note that the integration variable in the
decomposition~\eqref{eq:eEnu} should also be $\nu/c\tau$ for this
scaling analysis, which means that the relevant energy density per
mode should be $c\tau\varepsilon_{_E}(\nu)$.  This in addition should
be multiplied by the expanding length $c\tau$.  Therefore, the
dimensions of the three relevant quantities are
\begin{equation}
 [\nu/c\tau] = l^{-1} ,\qquad
 [V_\perp(c\tau)^2\varepsilon_{_E}(\nu)] = l^3 t^{-2} ,\qquad
 [\psi] = l^2 t^{-3} ,
\end{equation}
which are identical to the standard analysis of Kolmogorov's
spectrum.  Provided that $V_\perp(c\tau)^2\varepsilon_{_E}(\nu)$ is
expressed in a form of $(\nu/c\tau)^\alpha (\psi)^\beta$, we can fix
$\alpha$ and $\beta$ uniquely from the dimensions and thus expect the
Kolmogorov scaling, that is, $\tau^2\varepsilon_{_E}(\nu)\propto
(\nu/\tau)^{-5/3}$.  In fact we can confirm this scaling property
asymptotically from the right panel of
Fig.~\ref{fig:time-scale-long}.

\begin{figure}
 \includegraphics[width=0.49\textwidth]{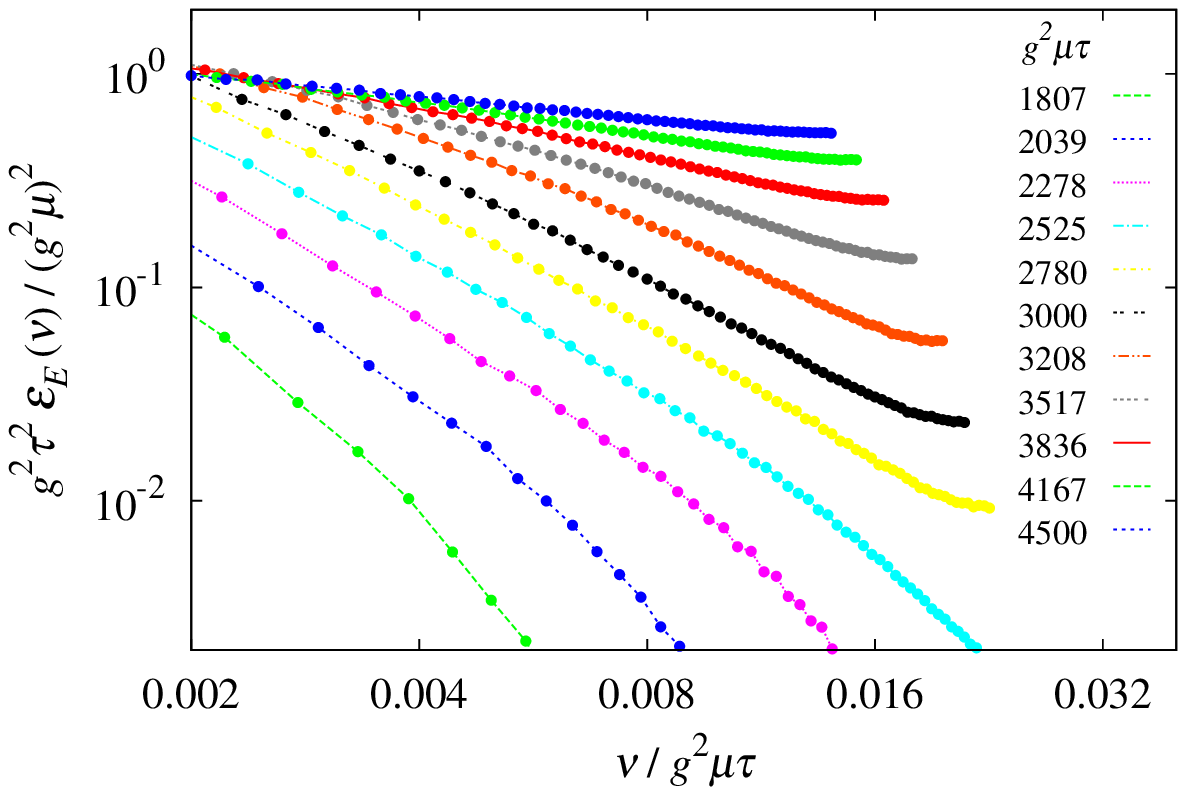}
 \raisebox{7pt}{\includegraphics[width=0.46\textwidth]{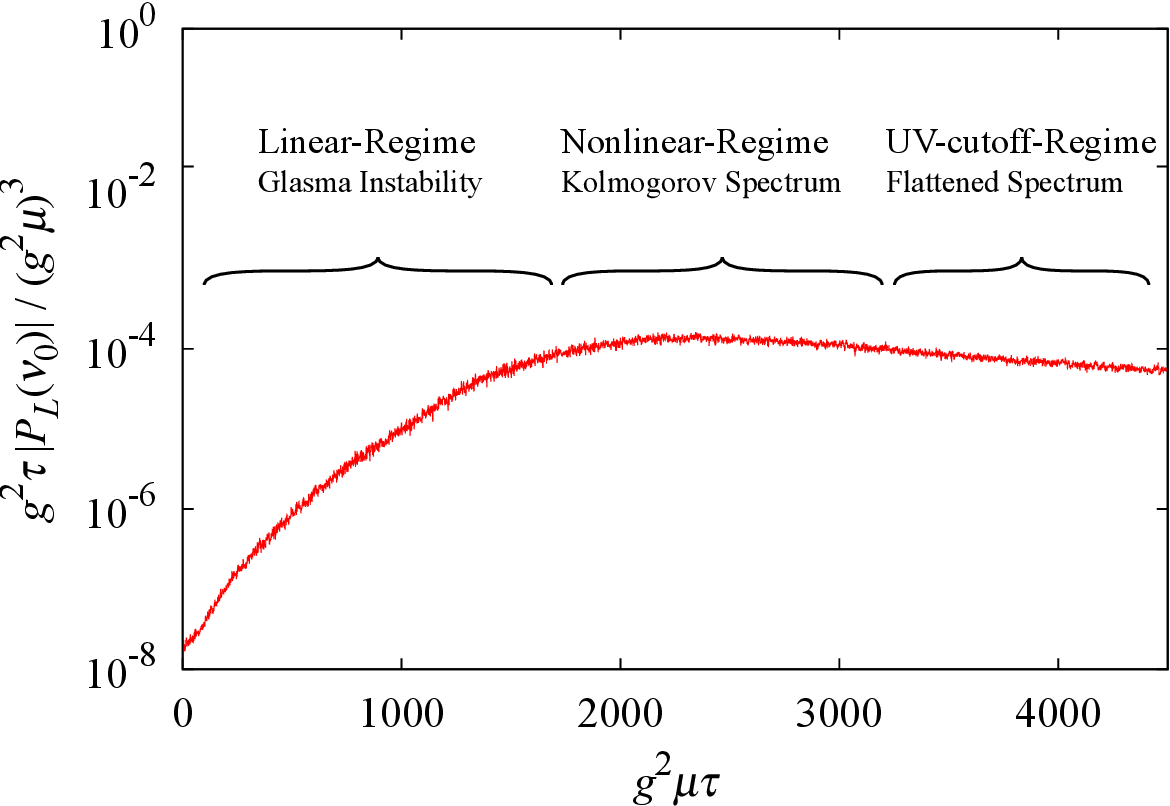}}
 \caption{Time evolution of the chromo-electric energy density as
   a function of the scaled longitudinal wave-number
   $\nu/(g^2\mu\tau)$ up to $g^2\mu\tau=4500$.\ \ Left: Results
   including later time spectra.  The horizontal axis is logarithmic
   to make it evident how the power-law spectrum spreads.\ \ Right:
   Longitudinal ``pressure'' with three regimes identified
   schematically.}
 \label{fig:stage}
\end{figure}

One might be wondering what the fate of the power-law spectrum should
be at even later times.  We understand from Fig.~\ref{fig:stage} that
the Glasma time evolution in the present MV-model simulation has three
stages;  the linear regime, the non-linear regime, and the UV-cutoff
regime.  In the linear regime the unstable modes grow up and the
non-linear effects are still minor.  When the amplitudes of $\nu\neq0$
modes become so large that the non-linearities can affect the time
evolution, the instability stops, as shown in the right panel of
Fig.~\ref{fig:stage}, and the non-Abelianization occurs
\cite{Arnold:2005vb,Rebhan:2004ur}.  Because non-linearities are no
longer negligible in this regime, the energy cascade due to
non-linearities is naturally expected, which would lead to the
Kolmogorov-type spectrum.  Eventually, as the turbulence decays into
higher UV modes, the UV-cutoff effect flattens the whole spectrum as
plotted in Fig.~\ref{fig:stage} up to $g^2\mu\tau=4500$.


To the best of our knowledge the present work is the first to exhibit
a Kolmogorov spectrum from the Glasma time evolution.  Our results
imply that the framework of the Glasma, in terms of classical
variables together with $\eta$-dependent fluctuations, may encompass a
rich physics contents including the QCD plasma instability.

Here, as a final remark in this section, we draw attention to related
works in analogous systems.  In the context of an ultra-cold gas, the
Kolmogorov spectrum has been obtained as a solution of the
Gross-Pitaevskii equation (GPE), which is a classical equation of
motion like the Glasma equations~\cite{Nowak:2010tm}.  The
calculational procedure by means of the GPE is similar to the Glasma
simulation, but in that case the microscopic mechanism of the energy
flow is understood in terms of the vortex dynamics.  In future studies
on the Glasma dynamics, it will be important to clarify the structure
of the energy flow, and it would be very interesting to see whether
the so-called non-thermal fixed point~\cite{Berges:2008wm} appears in
the strongly correlated Yang-Mills theory and Glasma.  Indeed, a
possible deviation in the power could be expected in Yang-Mills theory
by non-perturbative resummations~\cite{Carrington:2010sz}.


\section{Summary and Outlook}
\label{sec:summary}

In this paper we investigated the time evolution of the classical
equations of motion in the SU(2) pure Yang-Mills theory, especially in
the framework of the Glasma or the McLerran-Venugopalan model in (3+1)
dimensions.  Our emphasis has been put on the systematic
study of the instability behavior that occurs in fluctuations with
respect to the space-time rapidity $\eta$ when the initial conditions
break boost invariance.

Computing the longitudinal pressure component at non-zero wave-number
$\nu$, that is a conjugate of $\eta$, we have numerically confirmed
the Glasma instability that had been found in preceding
works~\cite{Romatschke:2005pm,Romatschke:2006nk}.  We carefully
checked the system size dependence of the instability to conclude that
the Glasma instability is a robust consequence from $\eta$-dependent
disturbances and is not sensitive to how the transverse and
longitudinal coordinates are discretized.  Hence, the Glasma
instability is certainly a physically well-defined phenomenon.  At the
same time, however, this means that taking the continuum limit does
not help with an earlier onset of the instability.  The typical time
scales we found for the Glasma instability is of order of hundreds in
terms of $g^2\mu\tau$, which is arguably too slow for being useful in
heavy ion collisions.  One should however keep in mind that the seeds
we have used in this numerical study are very small compared to the
Glasma fields, while in a realistic situation they are suppressed at most
by a power of the strong coupling constant.  There may also be a chance
that the non-linear effects could more or less accelerate the
instability.  We compared two results for the spectral distribution,
one from the superposition of individual calculations with the initial
condition that is given by a different wave-number $\nu_0$, and the
other from a single calculation with the initial condition that
contains all $\nu_0$'s.  We could observe the non-linear effects in
our numerical results, which would not affect the instability behavior
substantially, however, as long as the instability seed $\Delta$ is
reasonably small.

We have decomposed the energy density into Fourier modes to see how
the low wave-number modes cascade toward higher wave-number modes due
to non-linearity in the equations of motion.  Our motivation to do so
is to examine whether the so-called Kolmogorov's scaling can appear in
the turbulent system described by the Yang-Mills theory put in an
expanding geometry.  In contrast to the boost-invariant case, a
Kolmogorov-type scaling behavior in the energy spectrum a function of
the longitudinal wave-number $\nu$ may appear when the boost
invariance is broken by the initial fluctuations.  In this case, the
violation of boost invariance increases with time due to instability,
but the $\nu=0$ mode remains very large compared to the other
$\nu\neq0$ modes, which enables us to regard the $\nu=0$ mode as the
source for the energy injection.  Then, we could clearly observe that
the energy spectra approach an asymptotic form that shows the
Kolmogorov's scaling with the power $-5/3\simeq -1.67$.  As far as we
know, our work is the first example that exhibits the Kolmogorov-type
scaling law in the expanding systems.  Although it is difficult to
find any indication that the system indeed becomes isotropic and
thermalized in our simulation, the confirmation of the Kolmogorov-type
spectra is a promising signal for the general tendency heading for
thermalization.

There remain very interesting questions to be investigated in the
future.  In order to unveil the microscopic mechanism for the
instability, it would be very useful to make approximations on the
classical equations of motion in such a way that the analytical
treatment could be feasible.  For instance, as we have confirmed, the
instability mostly lies in the linear regime as long as its magnitude
specified by $\Delta$ is small enough.  Therefore the linearization of
the equations of motion in terms of fluctuating fields on top of the
boost-invariant CGC background should correctly describe the
instability in early times.  In this work we did not exploit such a
comparison between the numerical output and the analytical result from
the linearized equations.  This is because it is not quite efficient
to linearize the equations of motion given in terms of the lattice
variables.  The linearization is of course possible, but it would
eventually be reduced to similar calculations in the continuum
formulation.  Then, it would make much more sense to make a comparison
between the numerical and analytical results that are both formulated
in the continuum variables.  We have already executed test programs to
make sure if the numerical results from the Glasma simulations are
consistent with each other, given they are written in terms of lattice
and continuum variables.  We will report on this analysis elsewhere.

It is also an intriguing question to look for a link between
Kolmogorov's turbulent spectrum and the chaotic behavior generally
seen in the solution of the classical Yang-Mills equations.  This
problem may have some relevance for a deeper understanding of the
microscopic description of the entropy generating process and how the
thermalization is achieved at all.  The time evolution of the Glasma
and its physics contents deserve much more investigations in the
future.

\section*{Acknowledgments}
We would like to thank Kevin Dusling, Tuomas Lappi, and Raju
Venugopalan for discussions. This work has been supported in part by
JSPS Japan-France Integrated Action Program (SAKURA).  F.G.\ thanks
the Yukawa Institute for Theoretical Physics, where part of this work
has been done, for its hospitality and support.


\providecommand{\href}[2]{#2}\begingroup\raggedright\endgroup

\end{document}